\documentclass[12pt]{article}
\pdfoutput=1
\usepackage{amsmath,amssymb,amsfonts,bm}
\usepackage{enumerate}
\usepackage{graphicx}
\usepackage{accents}
\usepackage{setspace}
\usepackage{natbib}
\usepackage{color}
\DeclareMathOperator{\var}{Var}

\newcommand{\ev}{\mathbb{E}}

\title{Comparison of Cross-Validation Methods for Stochastic Block Models}%Goodness of Fit and Model Selection in Social Networks}
\author{Dabbs, Beau \\ \texttt{bdabbs@andrew.cmu.edu}
\and Junker, Brian \\ \texttt{brian@stat.cmu.edu}}
\begin{document}

\maketitle

\begin{abstract}
We introduce a novel cross-validation method that we call latinCV and we compare this method to other model selection methods using data generated from a stochastic block model.  Comparing latinCV to other cross-validation methods, we show that latinCV performs similarly to a method described in \cite{hoff2008modeling} and that latinCV has a significantly larger true model recovery accuracy than the NCV method of \cite{chen2014network}.  We also show that the reason for this discrepancy is related to the larger variance of the NCV estimate.  Comparing latinCV to alternative model selection methods, we show that latinCV performs better than information criteria AIC and BIC, as well as the community detection method infomap and a routine that attempts to maximize modularity.  The simulation study in this paper includes a range of network sizes and generative parameters for the stochastic block model that allow us to examine the relationship between model selection accuracy and the size and complexity of the model.  Overall, latinCV performs more accurate model selection, and avoids overfitting better than any of the other model selection methods considered.
\end{abstract}

\section{Introduction}

%Model selection is always an important endeavor for any researcher or statistician.  Recovering accurate models allows us to be more confident in our assertions based on those models.
%
%
%There are many methods that have been used for selection of network models, but many have them have never been compared to one another.
%\doublespacing

An increasing number of models have been developed for use with network data, many of which use latent variables to model the formation of ties in the network.%These models focus on understanding the nodes in the network, and using that information to predict tie formation.
Currently, few papers have analyzed the effectiveness of model selection methods for selecting among these network models.  In this paper we examine cross-validation methods for selecting among competing models, from both a theoretical and empirical point of view.  
Specifically we use cross-validation to estimate the risk for competing models when the true generating model is a stochastic block model (SBM) (\cite{Fienberg-Wasserman-sociometric-relations}, \cite{Holland-Lasky-Leinhardt-stochastic-blockmodels}).  We then compare the effectiveness of cross-validation with other model selection procedures.
%the use of cross-validation risk estimates as a model selection criterion for data generated from a Stochastic Block Model.  We compare various cross-validation methods to other model selection methods.    We use multiple performance criterion, including accuracy of true model recovery, to show that cross-validation methods are preferred for most generative parameters.

The cross-validation methods we consider evaluate the predictive risk of using a particular estimator to predict the value of ties in a new network.  For example, \cite{hoff2008modeling} uses a cross-validation method which assigns each each edge indicator in a network to a fold randomly in a V-fold cross-validation procedure.  Alternatively \cite{chen2014network} uses a method called network cross-validation (NCV) which first assigns each node to a fold, and then uses these nodal assignments to generate fold assignments for the edges in the network.  In this paper we also propose a third cross-validation method, latinCV, which attempts to balance the fold assignments across nodes while still directly assigning edges to folds.  We compare all three of these cross-validation methods to determine which is best at accurately selecting the true generative model.

We also consider the Bayesian information criterion (BIC), first introduced by \cite{schwarz1978estimating}.  BIC uses a penalized log-likelihood function to assess the goodness of fit of a given model.  The log-likelihood is typically maximized by the most complex model, and thus a penalty for complexity is used to balance goodness of fit and complexity of the model.  BIC is known to asymptotically choose the correct model in some settings, including multiple linear regression \citep{nishii1984asymptotic} and order estimation in Markov chains \citep{Csiszar-Shields-consistency-of-BIC}. But, we show that cross-validation provides a more accurate criterion for recovering the true model for pre-asymptotic network data.

%We also consider model selection methods that are specific to stochastic block models and the related problem of community detection.  First for choosing the number of blocks in an SBM we consider using the eigenvalues of the adjacency matrix as a model selection criterion.  It can be shown that a large gap should exist between the $K$th and $(K+1)$th eigenvalue when the true number of blocks is $K$ \citep{ng2002spectral}.  We thus use an \emph{eigenanalysis} method to choose the $K$ corresponding to this gap.

As an alternative to directly comparing the fits of SBMs, we also consider two community detection methods that attempt to detect an optimal set of community membership labels.  The first community detection method we consider is one that selects a partition of the nodes that maximizes modularity \citep{clauset2004finding}.  Modularity is a measure of the ratio of ties within communities to ties between communities that corrects for the degree of each node.  We refer to this method as modularity maximization.  The second community detection method we consider is known as the infomap method of \citet{rosvall2008maps}.  This method selects communities that allow for a minimum description length encoding of random walks on the networks and was found to be the best community detection method in \cite{lancichinetti2009community}.

In this paper we compare the performance of all of the model selection methods described above on data generated from stochastic block models with varying generative parameters.  In section \ref{sec:methods} we describe in detail all of the model selection and community detection methods.  In section \ref{sec:design} we review the design of our simulation study and the performance measures used to compare our methods.  We present the results of the simulation study in section \ref{sec:results} and discuss the importance of those results in section \ref{sec:discussion}.

%
%\singlespacing
%\begin{enumerate}
%\item Motivate Model Selection
%	\begin{itemize}
%	\item Also use accurate community membership recovery as an additional motivation, since we are considering NMI.
%	\end{itemize}
%\item Methods used for networks in general
%	\begin{itemize}
%	\item BIC (Example?)
%	\item CV - edge (used in \citet{hoff2008modeling}) and network \citep{chen2014network}
%	\end{itemize}
%\item Methods used to select communities (or among SBM)
%	\begin{itemize}
%	\item Eigenanalysis (Example?)
%	\item Recursive Bipartition \citep{bickel2015hypothesis} - Compared to networkCV in \citep{chen2014network}
%	\item Community Detection Methods
%		\begin{itemize}
%		\item Modularity Maximization (Find a standard way to do this for directed networks)
%		\item Infomap \citep{rosvall2008maps} (Shown to be best in \citet{lancichinetti2009community})
%		\end{itemize}
%	\end{itemize}
%\item We consider various cross-validation methods for network data.  We use the regime of Stochastic Block Models as an example.
%	\begin{itemize}
%	\item What method of cross-validation is preferred for Stochastic Block Models
%	\item How does CV compare to methods designed for community detection and SBM?
%	\end{itemize}
%\end{enumerate}
%
%
%\doublespacing

\section{Methods}
\label{sec:methods}
\subsection{Cross-Validation Methods}

Cross-validation methods divide the data into training and validation sets in order to estimate the performance of a given estimator when predicting new data.  These methods have been used for regression models with independent and identically distributed (i.i.d.) data as early as \cite{geisser1975predictive}.  \cite{arlot2010survey} provides an in depth review of both theoretical and empirical results of cross-validation methods for i.i.d. data.  In this paper we primarily examine empirical results for cross-validation on network data, where the edge indicators are only conditionally independent.  In section \ref{subsec:cv-variance} we examine some theoretical properties of our cross-validation estimators, but note that more work must be done to better understand the theoretical properties of cross-validation for networks.

The basic method of cross-validation requires us to divide the data, once or multiple times, into training and validation sets.  Model parameters are then estimated using the training set, and the model's performance is assessed on the validation set.  In this paper we will focus on $V$-fold cross-validation where each edge indicator, $Y_{ij}$, is assigned to one of $V$ folds.  We then estimate the tie probability, $\hat{p}^{CV}_{ij}$, for each edge using only the edge indicators not in the same fold as $Y_{ij}$.  Then, for a particular loss function $L$, we use the empirical loss between $\hat{p}^{CV}_{ij}$ and $Y_{ij}$ averaged over all of the edge indicators to estimate the predictive risk of the estimator, $\hat{p}$.  Formally we define the $VFCV$ estimate of risk using the following procedure:
%
%Before we begin describing the various cross-validation methods we will consider, let us set up some notation.  Since our dataset consists of an $n \times n$ matrix of observed edges, $Y$, we will assign each node pair $(i,j)$ with $1 \le i \ne j \le n$ to one of $V$ folds.  We represent this with an $n \times n$ matrix $A$ where $A_{ij} = v$ means that node pair $(i,j)$ is assigned to fold $v$.  To implement any cross-validation method we will do the following
{
%\singlespace
\begin{enumerate}
\item Partition the edge indicators, $Y_{ij}$, into $V$ disjoint \emph{folds}. We define a fold assignment matrix, $A$, where $A_{ij} = t$ iff $Y_{ij}$ is in fold $t$.%Assign each edge, $Y_{ij}$, a \emph{fold assignment} $A_{ij} \in \{1,...,V\}$.% in the network to one of $V$ folds.
\item For each fold, $t$% \in \{1,...,V\}$
\begin{itemize}
\item Define the \emph{training set} for fold $t$ to be $Y_{(t)} =  \{Y_{lk} : A_{lk} \ne t\}$
\item For each $Y_{ij}$ in fold $t$ ($A_{ij} = t$):
\begin{itemize}
\item Estimate $\hat{p}_{ij}^{CV} = \hat{p}_{ij} (Y_{(t)})$ 
\end{itemize}
\end{itemize}
\item Return the predictive risk estimate
$$
\widehat{R}_{VFCV}(Y,\hat{P}) = \frac{1}{n(n-1)}\sum_{t = 1}^{V} \sum_{Y_{ij}: A_{ij} = t}  L(Y_{ij}, \hat{p}_{ij}^{CV})
$$
\end{enumerate}
}
%
%{
%\singlespace
%\begin{enumerate}
%\item Generate a fold assignment matrix $A$
%\item For each model $M_1$,...,$M_T$ do:
%\begin{enumerate}
%\item For $v = 1$ to $V$ do:
%\begin{enumerate}
%\item Estimate parameters $\hat{\beta}_{v}^{CV}$ and latent variables $\hat{Z}_{v}^{CV}$ for model $M_t$ using only $Y^{-v} \defeq \{Y_{ij} : A_{ij} \ne v\}$
%\item Estimate $\hat{p}_{ij}^{CV,M_t} = \mathbb{P}(Y_{ij} = 1 | \hat{\beta}_{v}^{CV}, \hat{Z}_{v}^{CV})$ for each $i$ and $j$ with $A_{ij} = v$.
%\end{enumerate}
%\item Define $\hat{R}_{M_t}^{CV}(Y) = L(Y,\hat{P}^{CV,M_t}) = \frac{1}{n(n-1)} \sum_{i \ne j} L(Y_{ij},\hat{p}_{ij}^{CV,M_t})$
%\end{enumerate}
%\item Return $\hat{R}_{M_1}^{CV},...,\hat{R}_{M_{T}}^{CV}$
%\end{enumerate}
%}

To perform model selection, we then choose the model that minimizes $\hat{R}_{VFCV}(Y,\hat{P})$.  From this definition, we can see that we need to specify three things to define a $VFCV$ routine for networks: a loss function, $L$, the number of folds, $V$, and a method for assigning edge indicators to folds.  For this paper we focus on mean-squared error (MSE) as our loss function, defined to be
\begin{equation}
L_{MSE}(Y,\hat{P}) = \frac{1}{n(n-1)} \sum_{i \ne j} (Y_{ij} - \hat{p}_{ij})^2
\end{equation}
We focus on this loss function because its expected value decomposes directly into bias and variance as
\begin{equation}
\ev\left[L_{MSE}(Y,\hat{P})\right]  = \frac{1}{n(n-1)}\sum_{i \ne j} \left( \ev \left[\hat{p}_{ij}^K \right] - p_{ij} \right)^2 + \var \left[\hat{p}_{ij}^K \right] + \var \left[ Y_{ij} \right]. \label{mspe}
\end{equation}
%\begin{align*}
%\ev\left[L_{MSE}(Y,\hat{P})\right] & = \frac{1}{n(n-1)}\sum_{i \ne j} \left( \ev \left[\hat{p}_{ij}^K \right] - p_{ij} \right)^2 + \; \; \var \left[\hat{p}_{ij}^K \right] \qquad \quad + \var \left[ Y_{ij} \right] \\
%&  \qquad \qquad \qquad \qquad \left( \text{Bias} \right)^2 + \text{Estimation Variance} + \text{Data Variance}
%\end{align*}
This expected value is sometimes referred to as the mean squared predictive error, and this is the quantity we will attempt to estimate with cross-validation.  Mean-squared error is also a proper scoring rule, which means that the minimizer of the expect mean-squared error is always $\hat{p}_{ij} = p_{ij}$ for all $i$ and $j$ \citep{gneiting2007strictly}.  For the number of folds, we examine various choices empirically in the simulation study later in this paper.  We also consider three different fold assignment procedures which we define now.

\subsubsection{NCV}

First we define network cross-validation (NCV) proposed by \cite{chen2014network}.  To begin we assign each node to one of $V$ folds, $a_1,...,a_n \in \{1,...,V\}$.  Then, if $a_i = a_j = t$, we assign edge indicator, $Y_{ij}$, to fold $t$.  Otherwise, we assign $Y_{ij}$ to fold $0$.  Thus we have
\begin{equation}
A_{ij} = 	 \left\{
    \begin{array}{cc}
      v & L_i = L_j = v \\
      0 & L_i \ne L_j
    \end{array} \right.
\end{equation}
If an edge indicator is assigned to fold $0$, then that edge is present in the training set for every fold, but is never evaluated in the average used to estimate the predictive risk.  Figure \ref{base-folds} shows an NCV fold assignment when the nodes are ordered by their fold assignment, and figure \ref{random-folds} shows a randomized NCV fold assignment matrix.

The benefit of using the NCV method is that the training set for each fold is guaranteed to contain edge indicators for each node present in the training set, a property that we will refer to as \emph{node balance}.  However, one consequence of defining the folds this way is that the size of each training set is approximately $N \frac{V^2 - 1}{V^2}$, where $N$ is the total number of edges in the network, thus the training set sizes are comparable to those in $V^2$ fold random edge CV.  However the most important consequence of this assignment procedure is that the total number of edges used in validation sets is less than $\frac{N}{V}$, meaning the size of the validation set decreases as we increase $V$.

\subsubsection{latinCV}

In an attempt to both represent each node with edge indicators in each fold and use all of the edge indicators for evaluation of the model, we developed an alternative method we call \emph{latinCV}.  In experimental design, a latin square design can be used to balance treatment assignment across two factors so that each level of each factor is assigned once to each treatment condition.  If we think of the levels of the two factors as the rows and columns of a matrix, and the entries in the matrix as treatment assignments, a latin square is any matrix where each row and column has exactly one occurrence of each treatment.  For our fold assignment procedure, we treat the senders and receivers in our network as the two factors, and the folds as the treatment conditions.  Since we generally have more nodes than folds, we relax this definition of a latin square to be any matrix with equal occurrences of each fold assignment along each row and each column of the matrix.

\begin{figure}
\begin{center}
\includegraphics[width=2.1in]{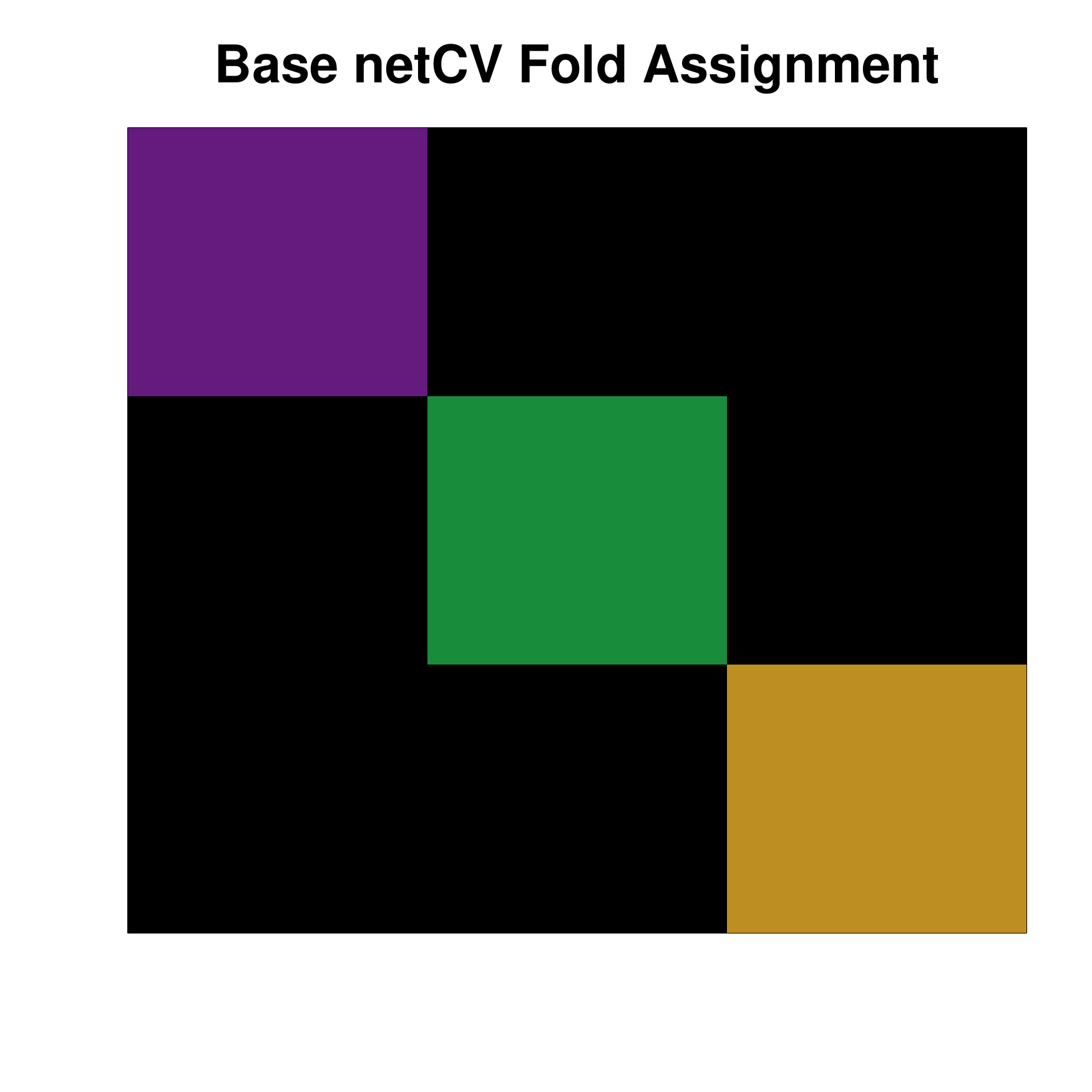}
\includegraphics[width=2.1in]{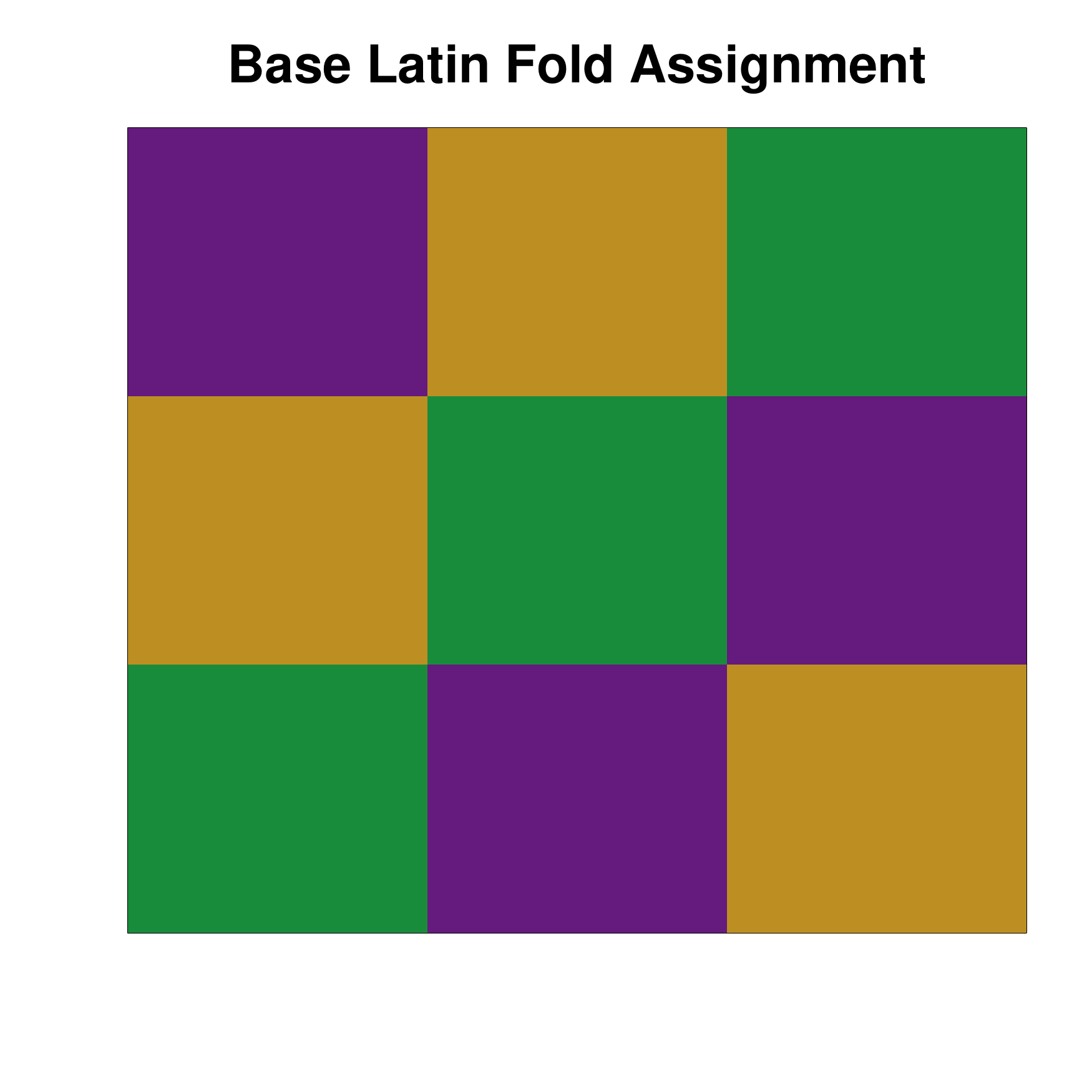}
\caption{Base 3 fold assignments for latinCV and netCV.  The colors purple, green, and gold correspond to folds 1-3.  Node-pairs colored black are never in test sets, but are always in training sets for netCV.}
\label{base-folds}
\end{center}
\end{figure}

To define our latinCV fold assignment procedure, we begin with a fixed fold assignment matrix, $A$, where each row and column of $A$ has an equal number of occurences of each fold.  A simple example of such a matrix can be seen in figure \ref{base-folds}.  We then permute then randomly, and 	independently permute the rows and columns of $A$ to obtain a random latinCV fold assignment matrix (figure \ref{random-folds}).
%
%The goal of latinCV, inspired by latin squares, is to guarantee that each row and each column of $A$ has an equal number of occurrences of each fold.  To generate such a matrix, we begin with a matrix that trivially satisfies this criterion.  Divide the matrix $A$ into $V^2$ blocks where each block is a $\frac{n}{V}$ by $\frac{n}{V}$ set of edge pairs. (See figure \ref{base-folds}.)  Thus we have our matrix represented as a $V$ by $V$ block matrix.  We then assign folds $\{1,2,...,V\}$ in order to the first row in the block matrix.  For the second row we assign folds $\{2,...,V,1\}$, and so on until for the $V$th row we assign the folds $\{V,1,...,V-1\}$.  To get a random fold assignment, we then independently reorder the rows of $A$ and then the columns of $A$.  Both of these reorderings preserve the balance among the rows and columns, resulting in our final fold assignment matrix which we will call $A_{latin}$.  Figure \ref{random-folds} shows an example of the randomized matrix.

\subsubsection{randomCV}

\begin{figure}
\begin{center}
\includegraphics[width=2.1in]{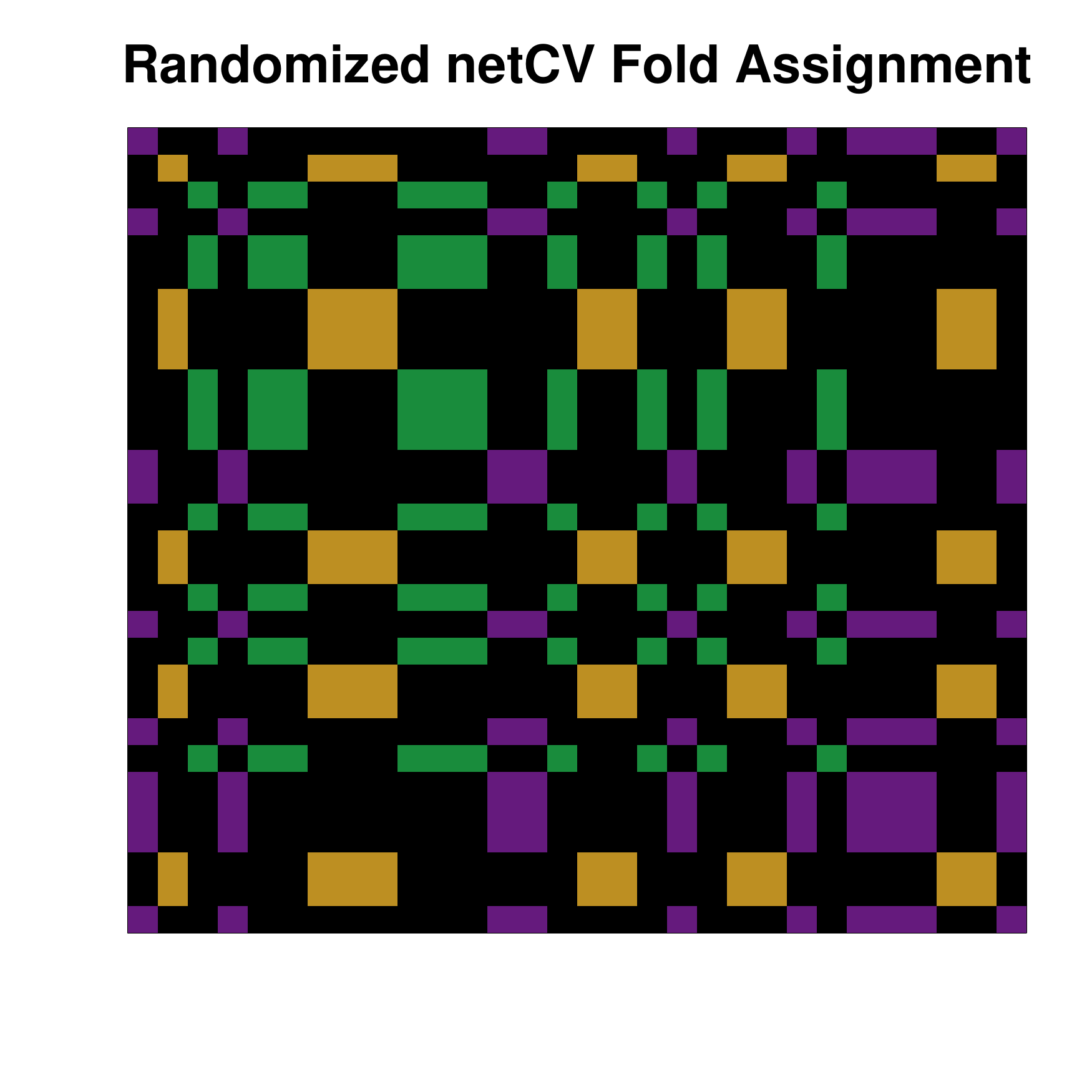}
\includegraphics[width=2.1in]{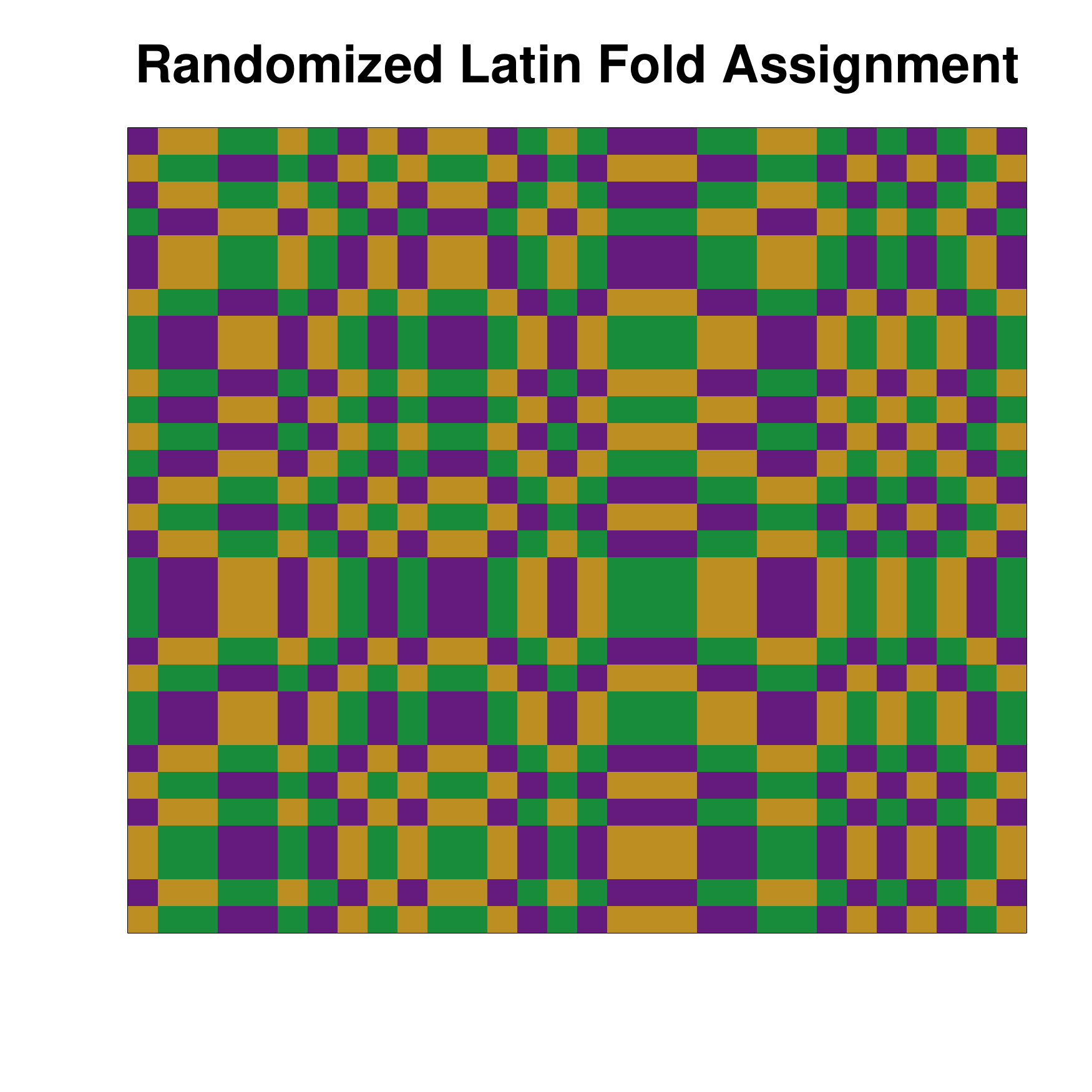}
\includegraphics[width=2.1in]{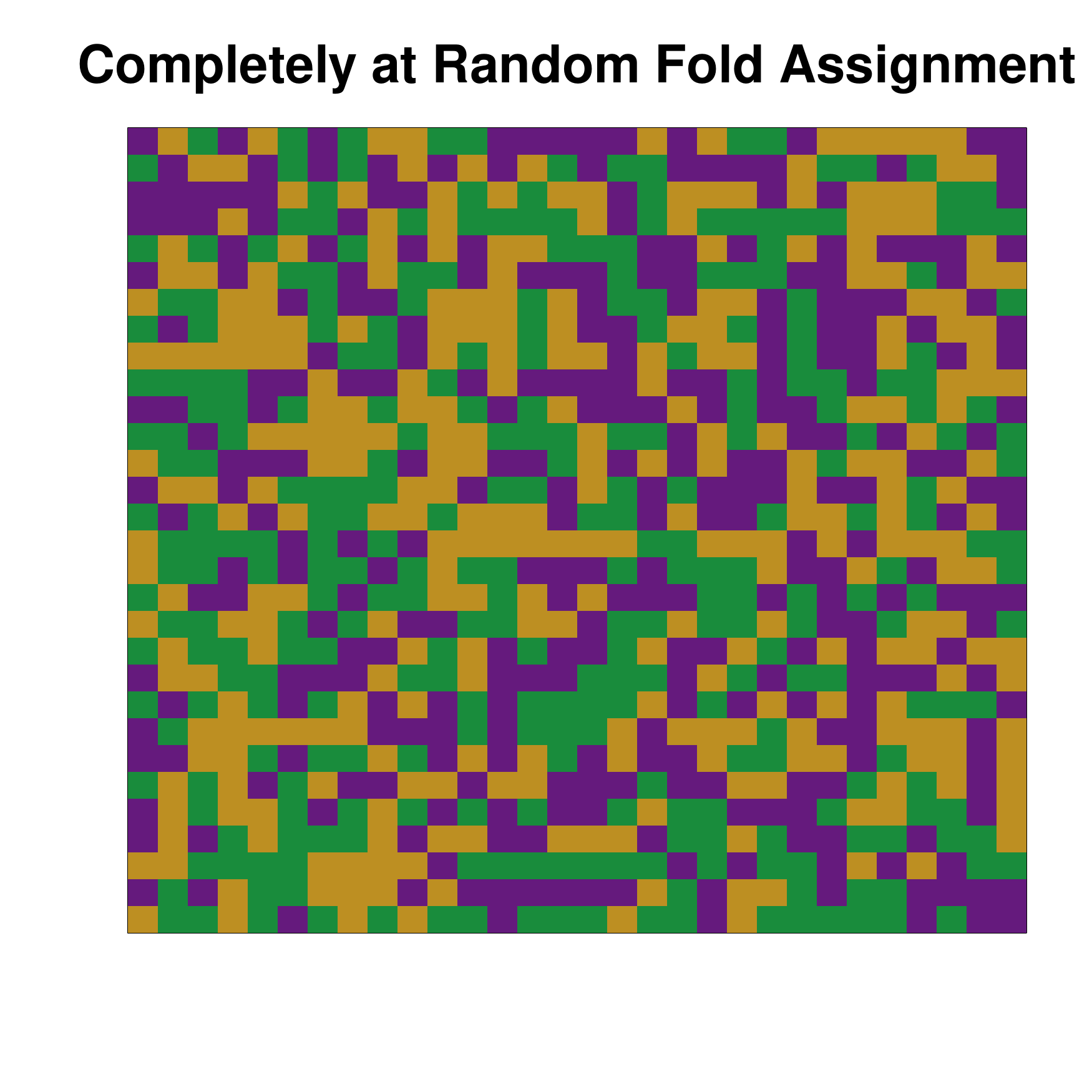}
\caption{The figure on the left shows a realization of a completely random fold assignment.  The center plot shows an example of latinCV fold assignment after randomizing each axis independently axes.  The right plot shows a netCV fold assignment after randomizing the axes together.}
\label{random-folds}
\end{center}
\end{figure}

The last cross-validation method we consider is the version presented in \cite{hoff2008modeling}, which we will refer to as randomCV throughout this paper.  This method is the simplest procedure, and does not guarantee balanced representation of each node in each fold.  To obtain a fold assignment matrix for randomCV, we simply select uniformly at random among all possible fold assignment matrices with an equal number of occurences of each fold.  Thus, we guarantee that each fold is the same size, but do not guarantee the node balance of latinCV and NCV.  An example of a randomCV fold assignment matrix can be seen in figure \ref{random-folds}.

%
%\begin{itemize}
%\item Description of networkCV method.
%\item Comparison of $K$-fold network CV to $K^2$-fold random CV
%\end{itemize}

\subsection{Information Criteria}
\label{ic-methods}
We also consider using information criteria as model selection methods.  We primarily examine the use of the Bayesian information criterion and the Akaike information criterion (AIC) \cite{akaike1998information}.  Both of these methods use a criterion that involves the log-likelihood and a penalty for the complexity of the model being used.  A penalty is required because without it more complex models will generally be able to capitalize on chance and fit the current dataset well, while generalizing poorly.  For nested models where MLEs are used to estimate parameters the log-likelihood is guaranteed to increase, and thus the most complex model will always be chosen without something to control the model complexity.  For both AIC and BIC, we select the model that minimizes the criterion.

\subsubsection{Akaike Information Criterion}

We define the Akaike Information Criterion to be
\begin{equation}
AIC = -2 \log \left[ \mathbb{P}(Y = y | \hat{B},\hat{\pi}, \hat{\gamma}) \right] + 2 d.
\end{equation}
The penalty term for AIC, involves the degrees of freedom of the model being estimated, $d$.  For regression models, $d$ is simply the number of coefficients that need to be estimated.  For many network models, including the stochastic block model, the exact value of $d$ is unclear, and we discuss the possible values for $d$ in section \ref{model-def}.

\subsubsection{Bayesian Information Criterion}

The Bayesian Information Criterion also penalizes based on the degrees of freedom in the model, but penalizes an amount proportional to the sample size.  We define BIC here to be
\begin{equation}
BIC = -2 \log \left[ \mathbb{P}(Y = y | \hat{B},\hat{\pi}, \hat{\gamma})\right] + d \log(n(n-1)).
\end{equation}
For undirected networks, the total number of possible edges in the network is $n(n-1)$, and this is the sample size we use for BIC.  Again, we save our discussion about the particular value of $d$ for section \ref{model-def}.  We note here that BIC will always prefer smaller models than $AIC$.  For any network with just for or more nodes, $n$, we have $\log(n(n-1)) > 2$, and thus BIC penalty for complexity is greater even for relatively small networks.  For this reason, BIC will frequently prefer much smaller models as the number of nodes in the network grows.

\subsection{Community Detection Methods}
\label{subsec:cd-methods}

The model selection methods described above are relatively general, and can be applied to models more general than the stochastic block model.  In this section we focus on methods designed to detect communities within a network, and thus are apparently more suited to block models, and we can compare the number and accuracy of the communities found by these methods, with the generating SBM.

\subsubsection{Modularity Maximization}
\label{subsubsec:modularity}

Modularity is a measure of the quality of a division of a network into clusters, first introduced in \cite{newman2004finding} for undirected networks.  Modularity was extended to directed networks in \cite{leicht2008community}, and we use their definition of \emph{directed modularity} to be
\begin{equation}
Q = \sum_{i \ne j} \left(\frac{Y_{ij}}{m} - \frac{k_{i}^{out}}{m} \frac{k_{j}^{in}}{m} \right) \delta_{\pi_i,\pi_j} = \frac{1}{m} \sum_{i \ne j} \left(Y_{ij} - \frac{k_{i}^{out} k_{j}^{in}}{m}  \right) \delta_{\pi_i,\pi_j}.
\end{equation}

In the above equation $m = \sum_{i \ne j} Y_{ij}$ is the total number of edges in the network, and $\delta_{l,k}$ is the Kronecker delta symbol taking the value of $1$ when $l = k$, and $0$ otherwise.  Modularity is a measure of how more connected communities are than would be expected on average once we correct for the degree of the nodes within each community.  Thus modularity provides a measure of the quality of any community assignment for the network.  Unfortunately, finding the set of community assignments that maximizes modularity is an NP-complete problem \citep{brandes2008modularity}.  Due to this complexity, we use a greedy algorithm proposed by \cite{leicht2008community} to choose a set of community assignments that approximately maximizes modularity.

\subsubsection{Infomap}

The next method we consider has its roots in information theory.  Suppose that we have a random walker who begins at an arbitrary node in the network and then walks to an adjacent node selected uniformly at random among all adjacent nodes.  We want to create a code, a set of $1$s and $0$s, that can describe any random walk with a minimum length.  We call this code a \emph{Minimum Description Length} (MDL) code.  If the communities are densely connected within the community, with many fewer connections between communities, \cite{rosvall2008maps} showed that you can use codes to represent communities, and decrease the description length for all steps taken within a community, leading to a large reduction in description length.

Similar to the modularity maximization methods, it is computationally infeasible to consider all of the possible community assignments and choose the one with minimum description length.  \cite{rosvall2008maps} describe a greedy algorithm that begins with each node belonging to its own community, and then iteratively combines the two communities which give the greatest decrease in description length.  
Once a set of communities for which no combination of two communities will decrease the description length is found, a simulated annealing approach is used to search the space of community memberships near the greedy optimum, resulting in a set of communities with marginally better MDL.  For this paper we use the \texttt{cluster\_infomap} method in the \texttt{igraph} package \citep{package:igraph} which implements the infomap method.

\subsection{Stochastic Block Model}

\subsubsection{Model Definition}
\label{model-def}

The stochastic block model is a model for network data that formalizes the idea that each node in a network belongs to a community, and that each node makes ties with other nodes based on their communities.  We represent our data as a matrix of edge indicators, $Y_{ij}$.  The full matrix, $Y = (Y_{ij})$, is an $n \times n$ matrix, where $n$ is the number of nodes in the network.  For this paper we assume that the network is directed, and thus $Y_{ij}$ is not necessarily equal to $Y_{ji}$.  We also assume that no self-ties are possible, i.e. $Y_{ii} = 0$.

The stochastic block model first supposes that each person has a latent block membership, $\pi_i \in \{1,...,K\}$, which we do not observe.  We then define a block tie probability matrix, $B = (b_{lk})$, where $b_{lk}$ is the probability of observing an edge between a node in block $l$ to a node in block $k$.  Thus, conditional on $\pi_i$, $\pi_j$, and $B$, we have
\begin{equation}
\mathbb{P}(Y_{ij} = 1 | \pi_i, \pi_j, B) = b_{\pi_i,\pi_j}.
\end{equation}
Thus the complete log-likelihood of the block memberships and block tie probability matrix is
\begin{equation}
\ell(\hat{B}, \hat{\Pi} ; y) = \sum_{i \ne j} \log \left( \mathbb{P}(Y_{ij} = y_{ij} | \hat{B}, \hat{\Pi} ) \right) = \sum_{i \ne j} Y_{ij} \log(b_{\pi_i,\pi_j}) + (1 - Y_{ij}) \log(b_{\pi_i,\pi_j}).
\end{equation}
Since the $\pi_i$ are modeled as latent variable, we assume that they have some prior distribution
\begin{equation}
\pi_i \sim multinomial(\gamma).
\end{equation}
The parameter, $\gamma$, is constrained to be a length $K$ vector that sums to $1$, and each entry, $\gamma_l$ is the probability of a node belonging to block $l$.

In order to estimate the parameters in a SBM we first fix the total number of blocks, $K$, and then estimate both $\hat{\Pi}$ and $\hat{B}$.  Unfortunately, finding the set of block assignments that maximizes the likelihood is a combinatorically difficult problem, and thus we must use methods such as spectral clustering or expectation maximization to approximate the maximum likelihood estimators for $\hat{\Pi}$.  In the following sections we review the methods for estimating both $\hat{B}$ and $\hat{\Pi}$.

\subsubsection{Model Estimation for SBM}

When fitting the Stochastic Block Model to network data we need to assign each node to one of the $K$ blocks and then we need to estimate the block tie probability matrix, $B$.  Given a set of block membership assignments, $\hat{\Pi}$, the maximum likelihood estimates for $\hat{B}$ are relatively straightforward.  If we consider the log-likelihood of $B$ conditional on the block memberships, $\hat{\Pi}$, we have
\begin{equation}
\log \left( \mathbb{P}(Y | \Pi = \hat{\Pi}, B) \right) = \sum_{l,k} Y_{(lk)} \log(b_{lk}) + (n_{lk} - Y_{(lk)}) \log(1 - b_{lk}),
\end{equation}
where $Y_{(lk)} = \sum_{i \ne j} Y_{ij} \mathbb{I} \{ \pi_i = l \& \pi_j = k\}$ is the total number of edges observed between block $l$ and $k$, and $n_{lk}$ is the total number of edges possible between blocks those blocks.  From this conditional log-likelihood it's easy to see that the maximum likelihood estimate for each $b_{lk}$ is simply
\begin{equation}
\hat{b}_{lk} = \frac{Y_{(lk)}}{n_{lk}}.
\end{equation}
Thus, if we use any community detection algorithm to estimate $\hat{\Pi}$, we can obtain an estimate of $\hat{B}$ using the MLE conditional on that set of community assignments.

\subsubsection{Spectral Clustering SBM Estimates}

For this paper we use the spectral clustering method described in \cite{chen2014network}.  Consider the singular value decomposition of the adjacency matrix, $Y = U \Sigma V^T$.  Here $\Sigma$ is the diagonal matrix containing the singular values $\sigma_1 \ge \sigma_2 \ge \cdots \ge \sigma_n$.  If our goal is to assign each node to one of $K$ communities, then we consider the submatrix $U^K \in \mathbb{R}^{n \times K}$ corresponding to the leading $K$ right singular vecotrs.  The rows of $U^K$ then represent coordinates in $\mathbb{R}^K$ for each node in the network, and we can implement a standard clustering algorithm on these points.  Here we use a version of $K$-means clustering implemented in the \texttt{stats} package of \texttt{R} to determine cluster assignments.

It turns out that this relatively simple algorithm is a consistent estimator of the community memberships as we increase the number of nodes in the network to infinity when the network is undirected \citep{lei2015consistency}.  The intuition behind this result is that the underlying tie probability matrix, $P$, is guaranteed to be of rank $K$ for a stochastic block model, and the observed network, $Y$, can be thought of as a noisy version of $P$.  Performing $K$-means clustering on the $K$ leading right singular vectors of $P$ would result in a perfect recovery of the block memberships since there are $K$ unique rows in the matrix of $K$ leading singular vectors for $P$, one for each of the $K$ blocks in the network.  Thus, if we instead perform $K$-means clustering on the $K$ leading singular vectors of $Y$ we still expect relatively strong separation of the communities.  In this paper we use this spectral clustering algorithm as an initialization routine for expectation maximization.

\subsubsection{Variational EM}
After obtaining an initial set of block memberships, either using spectral clustering or sampling from a prior distribution over block memberhips, we use a variational expectation-maximization (EM) algorithm to iteratively update our estimates of $\hat{B}$ and $\hat{\pi}$ until $\hat{B}$ converges to a fixed point.  We use a version of the EM algorithm based on the mean field approximation described in \cite{zhang2012comparative}, but modified to work for directed networks.  A full description of this algorithm can be obtained by request from the author.  Experimentation with the spectral clustering method and the EM algorithm found that the EM algorithm frequently converges in less than 2 steps from the spectral clustering start, but resulted in more accurate recovery of the block memberships than spectral clustering alone.  For example, for 60 node networks generated from a 2 block SBM we found that EM with spectral clustering was able to recover the correct block assignments $99.8\%$ of the time, whereas spectral clustering alone recovered the correct assignments only $69.8\%$ of the time.  

We also considered initializing the EM algorithm with multiple random initial membership vectors.  For the same set of 60 node 2 block networks above EM with 100 random starts was only able to recover correct block assignments $0.7\%$ of the time.  Thus EM with SC initialization provided noticeably better performance compared to just SC and also outperformed multiple random initializations.

\subsubsection{Degrees of Freedom in SBM}

We mentioned before that the degrees of freedom used for both AIC and BIC is unclear for models such as the stochastic block model.  The primary issue is determining the degrees of freedom contributed by estimating the latent variable, $\hat{\Pi}$.  If we knew the block assignments, $\Pi$, then estimating the $K^2$ block tie probabilities in $B$ would contribute $K^2$ degrees of freedom.  However, since we are estimating $\pi$, this estimation adds additional degrees of freedom.  

%There are some examples in the literature where the contribution to the degrees of freedom due to estimating the latent classes is assumed to be $0$ \citep{daudin2008mixture}.  However, we found this penalty to be too conservative.  
For the mixed membership stochastic block model, which allows the block memberships to be partial, \cite{airoldi2008mmsb} used a BIC criterion that counted the parameters to be estimated in both $\hat{B}$ and in the Dirichlet prior for the memberships, $\hat{\alpha}$, for a total of $d = K^2 + K$ degrees of freedom  In our case, the prior distribution over the block memberships is multinomial, and the prior, $\gamma$, is restricted to sum to one.  So, we suppose here that $K-1$ degrees of freedom are contributed for the estimation of the latent block assignments, and thus using $d = K^2 + K - 1$.

For BIC, it has also been suggested that the effective sample size should be less than the number of edge indicators in the network, $n(n-1)$ \citep{airoldi2007combining}.  However, the reason for this reduction in sample size is due to the way networks are frequently sampled, where we observe the presence of edges, but do not necessarily observe the absence of edge.  For this reason it was suggested to only count the observed edges, those where $Y_{ij} = 1$, in the sample size.  For our simulation study, however, we know that we are observing directly both present and missing edges, and thus can count all possible edges as part of the sample size.

\section{Simulation Study Design}
\label{sec:design}

%
%\textbf{Potential Plots:}
%\begin{itemize}
%\item Heatmap of Performance, as predicted by a model on the simulated output
%\end{itemize}
%\subsection{Simulation Study Design}

Stochastic block models can be used to model the community structure that can be observed in networks in fields from biology to sociology.  In this section we describe our choices for simulation parameters, and define our performance criteria for comparing the competing model selection methods.

\subsection{Simulation Study Parameters}

\begin{figure}
\begin{center}
\includegraphics[width=6.5in]{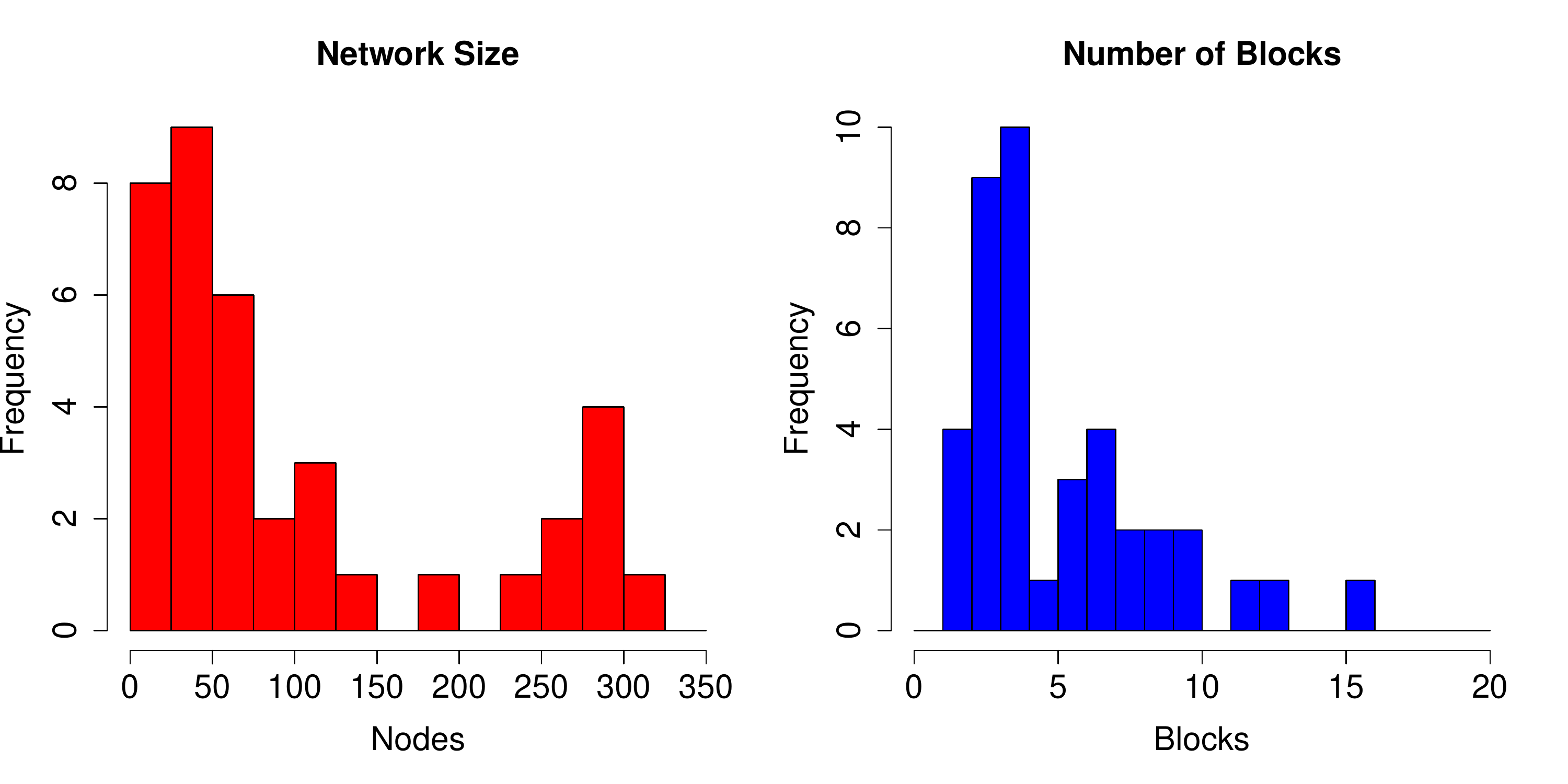}
\caption{\textbf{Left:} Histogram of the number of nodes in a selection of networks drawn primarily from the social science and medical literature.  There were additionally 4 networks with 500-2000 nodes and 4 networks with greater than 200,000 nodes in our sample of networks from the literature. \textbf{Right:} Histogram of the number of blocks found in the same selection of networks.  There were additionally 3 networks with greater than 5000 blocks, and one network with 40 blocks in our sample.}
\label{hist-1}
\end{center}
\end{figure}

When simulating data for any experiment one important consideration is the sample size.  For network data the sample size, $n$, is simply the number of nodes in the network.  To determine a range of node sizes from which to simulate networks we surveyed the social network literature \citep{dabbs-2016-lit-review}.  	Figure \ref{hist-1} shows a histogram of the number of nodes found from that survey.  Most of the networks contained fewer than 300 nodes, with a few networks having more than 1000 nodes.  For our simulation study we will generate networks with $n = 30, 60, 120$, and 300 nodes.

While surveying the literature, we also kept track of the number of blocks either known to be present or used to fit a model.  A large majority of the networks had ten or fewer blocks, and thus we will simulate data from SBMs with $K = 1, ..., 10$ blocks.  For the smaller networks with $120$ or fewer nodes, we only simulated networks with up to $5$ blocks.  Even when using the entire observed network, estimation of the block memberships was difficult for more than $5$ blocks when there were too few edge indicators.  This difficulty caused all model selection methods to perform poorly, frequently selecting smaller, less complex models.

%Once we know the number of nodes and the number of blocks from which to simulate networks, we still need to know two more specifications.
We also need to know how to assign memberships, $\pi_i$, for each node $i$.  We start by considering membership assignments that assign an equal number of nodes to each of the $K$ possible blocks.  Then we consider assignments that result in a power law distribution on the block sizes.  In particular, for a $K$ block model, we drew $K$ samples from a pareto distribution with a shape parameter, $\alpha = 1.5$, and a minimum value of $x_{min} = \frac{n}{3K} \cdot $.  These values were chosen so that the expected value of each block size was finite and equal to $\frac{n}{K}$.  We then ordered the block sizes, and used the average size of the largest block to the smallest block.  This process generally resulted in networks with one large block, and a few smaller blocks decreasing slowly in size.  For example, for the networks with $120$ nodes and $5$ blocks the number of nodes in each block were $(64,22,14,11,9)$.

Lastly we need to know the probability of there being a tie between a node in block $l$ and another node in block $k$ for all $l,k = 1, ..., K$.  This set of \emph{block tie probabilities} is generally represented using a matrix $B$ where the entries are each $b_{lk} = \mathbb{P}(Y_{ij} = 1 | \pi_i = l, \pi_j = k)$.  Many examples where an SBM seems appropriate are those where each block corresponds to a community of nodes that are more connected with each other than with those outside of the community.  For this reason we consider a set of tie probability matrices where the probability of a tie between nodes in different blocks is fixed at a constant, $b$, and the probability of a tie between nodes in the same block is $rb$ for some $r > 1$.  Networks generated with these types of tie probability matrices also conform to the assumptions of community detection methods such as modularity maximization reviewed in section \ref{subsec:cd-methods}.

We chose ranges for $b$ and $r$ that corresponded to a set of relatively difficult model selection problems.  We had $b$ range from $0.01$ to $0.05$ to $0.1$, and we had $r$ vary among $3$, $4$, and $5$.  While it is not uncommon to see networks where the tie density ratio, $r$, is 10 or greater, for larger values of $r$ most model selection methods perform relatively well, with near perfect model selection accuracy.  By choosing a range that includes more difficult model selection tasks, we can better distinguish the abilities of our model selection methods.  Our complete set of data generation parameters is summarized as follows:

%We also consider tie probability matrices where each element $b_{lk} \sim Uniform(0,1)$.  While tie probability matrices of this form are rare, some datasets have been found that have tie densities that have no obvious form - see for example the network of neural connections in C. Elegans \citep{pavlovic2014stochastic}.
%\singlespace
\begin{itemize}
\setlength\itemsep{0em}
\item Number of Nodes - 30, 60, 120, 300
\item Number of Blocks - 1, 2, 3, 4, 5
\item Block Membership Distribution:
\begin{itemize}
\setlength\itemsep{0em}
\item Equal Sized Blocks
\item Block Sizes with Power Law Proportions
\end{itemize}

\item Tie Probability Matrix -
\begin{itemize}
\setlength\itemsep{0em}
\item Densely Connected Communities - $b = .01, .05, .1$ ; $r = 3, 4, 5$
	\begin{equation}
	B = b \cdot \left( \begin{array}{cccc} r& 1 & \cdots & 1 \\ 1 & r & \cdots & 1 \\ \vdots & \cdots & \ddots & 1 \\ 1 & \cdots & 1 & r     \end{array} \right)
	\end{equation}
\end{itemize}
%	\begin{itemize}
%		\item Sparsity - $B = r \cdot B_0$
%		\item Tie Probability Differences $| b_{lk} - b_{ij} |$
%	\end{itemize}
%\item - Total Cells - 1100
\end{itemize}
%\doublespace

For each cell in the experimental design, any combination of $(n,K,b,r)$ and the particular method for assigning nodes to blocks we sampled $312$ independent networks with the specified generative distribution.  We then used all of the methods described in section \ref{sec:methods} to estimate the number of blocks, nodal block assignments, and the block tie probability matrix.  For the cross-validation methods, we also varied the number of folds used to be one of $V = 3, 5,$ or $10$.

%
%\subsection{Model Estimation}
%
%
%Since we are simulating data from stochastic block models with 1 to 10 blocks, we chose to estimate model parameters for SBMs with 1 to 11 blocks. This set of estimating models allows the opportunity to choose more than the true number of blocks for each $K$ in the simulation study.  To estimate parameters for the Stochastic Block Model we will be using a variational EM algorithm based on the mean-field approximation \citep{em-mean-field}.  To initialize the algorithm we use spectral clustering on the adjacency matrix to obtain an initial set of block memberships, similar to the method in \citet{chen2014network}.  
%
%Since we are using a cross-validation procedure to estimate risk for these models, our estimation methods need to be able to deal with missing data.  The EM algorithm can deal with missing data by simply computing the conditional likelihood marginalized over the missing ties.  Due to the conditional independence of the ties, marginalizing over any single tie is equivalent to computing the likelihood without that tie.  Unfortunately, spectral clustering requires a complete dissimilarity matrix to perform the singular value decomposition.  Since we are only using spectral clustering to initialize our algorithm, we simply complete our matrix by substituting the mean tie value for each missing element in the matrix, including the diagonal entries.  We also considered substituting missing values with $0$'s, but found that using the average tie-value resulted in better recovery of the block memberships.

\subsection{Performance Measures}

We considered both model selection accuracy and mean squared error (MSE) loss in estimating tie probabilities.  We define model selection accuracy to be the percentage of times a model selection method chose the estimator with $K$ blocks when the data was truly generated from a $K$ block SBM.  %We use this measure because our primary goal is to determine which methods are best for model selection, with the goal of eventually using these methods to perform more complex model selection tasks.

We define the mean squared error loss to be
\begin{equation}
L_{MSE}(P,\hat{P}) = \frac{1}{n(n-1)} \sum_{i = 1}^n \sum_{j \ne i}^n (p_{ij} - \hat{p}_{ij})^2.
\end{equation}
We use this criterion as a measure of the overall goodness of fit to the true tie probabilities underlying the observed network.  We also note that this mean squared error is directly related to the mean squared predictive error defined in equation \ref{mspe}, by the following equation
\begin{equation}
\ev[L_{MSE}(Y,\hat{P})] = \ev[L_{MSE}(P,\hat{P})] + \sum_{i \ne j} \var[Y_{ij}]
\end{equation}
Since cross-validation attempts to estimate mean squared predictive error, and then chooses the model that minimizes this error, cross-validation methods should also select models that minimize our MSE criterion, $L_{MSE}(P,\hat{P})$.  For this reason, MSE acts as both a check of overall goodness of fit, as well as a way of testing how well cross-validation methods are able to minimize the criterion they attempt to estimate.

%As mentioned above, the primary goal of this simulation study is determine which methods are most accurate at recovering the true model.  Preliminary simulations suggest that cross-validation performs remarkably well at the task for stochastic block models.  When using NMI as the performance criterion, previous work by \citet{lancichinetti2009community} showed that the infomap method performed best.  However, their studied used a different set of simulation parameters related to the GN bencmarks \citep{ng-benchmarks}.  Finally, for the task of minimizing the loss between the estimated and true tie probability matrices, we expect the cross-validation methods to perform best.  Since cross-validation provides an estimate of risk (the expected loss between $P$ and $\hat{P}$), and we choose the model that minimizes the CV estimate of risk, we would expect the loss between that estimate and $P$ to be relatively small.

\section{Results}
\label{sec:results}

In this section we examine the performance of the model selection methods in the simulation study.  Table \ref{overall-accuracy} shows the average model selection accuracy over all of the generation parameters, along with a 95\% confidence interval based on the standard error of the estimates of the average accuracy.  latinCV and randomCV have significantly larger accuracy than any of the other methods considered.  latinCV does perform slightly better than randomCV, but their difference is not statistically significant.  In this section we compare the performance of these methods as we vary some of the network simulation parameters, and try to understand when these methods perform well and when they perform poorly.

%\begin{itemize}
%\item the CV methods perform best averaged over the entire simulation study.
%\item log-likelihood performs the worst, always selecting models that are too large.
%\end{itemize}
%
% latex table generated in R 3.2.2 by xtable 1.8-2 package
% Thu Apr 28 10:44:38 2016
\begin{table}[ht]
\centering
\begin{tabular}{|l|rrr|}
  \hline
  \multicolumn{4}{|c|}{Model Selection Accuracy} \\ \hline
Method & Average & .025\% & .975\% \\ 
  \hline
latinCV & 0.422 & 0.417 & 0.426 \\ 
  randomCV & 0.421 & 0.417 & 0.426 \\ 
  NCV & 0.383 & 0.379 & 0.388 \\ 
  BIC & 0.380 & 0.376 & 0.385 \\ 
  modularity & 0.307 & 0.303 & 0.311 \\ 
  infomap & 0.203 & 0.200 & 0.206 \\ 
  AIC & 0.031 & 0.030 & 0.033 \\ 
  log-likelihood & 0.000 & 0.000 & 0.000 \\ 
   \hline
\end{tabular}
\caption{Average model selection accuracy over all network simulation parameters.  We compare 10 fold latinCV and randomCV to 3 fold NCV.  These were the optimal number of folds for each method.}
\label{overall-accuracy}
\end{table}

\subsection{Comparing Cross-validation Methods}
%
%\begin{itemize}
%\item First we compare number of folds, then we compare the methods side by side.
%\end{itemize}

\begin{figure}
\begin{center}
\includegraphics[width=4.0in]{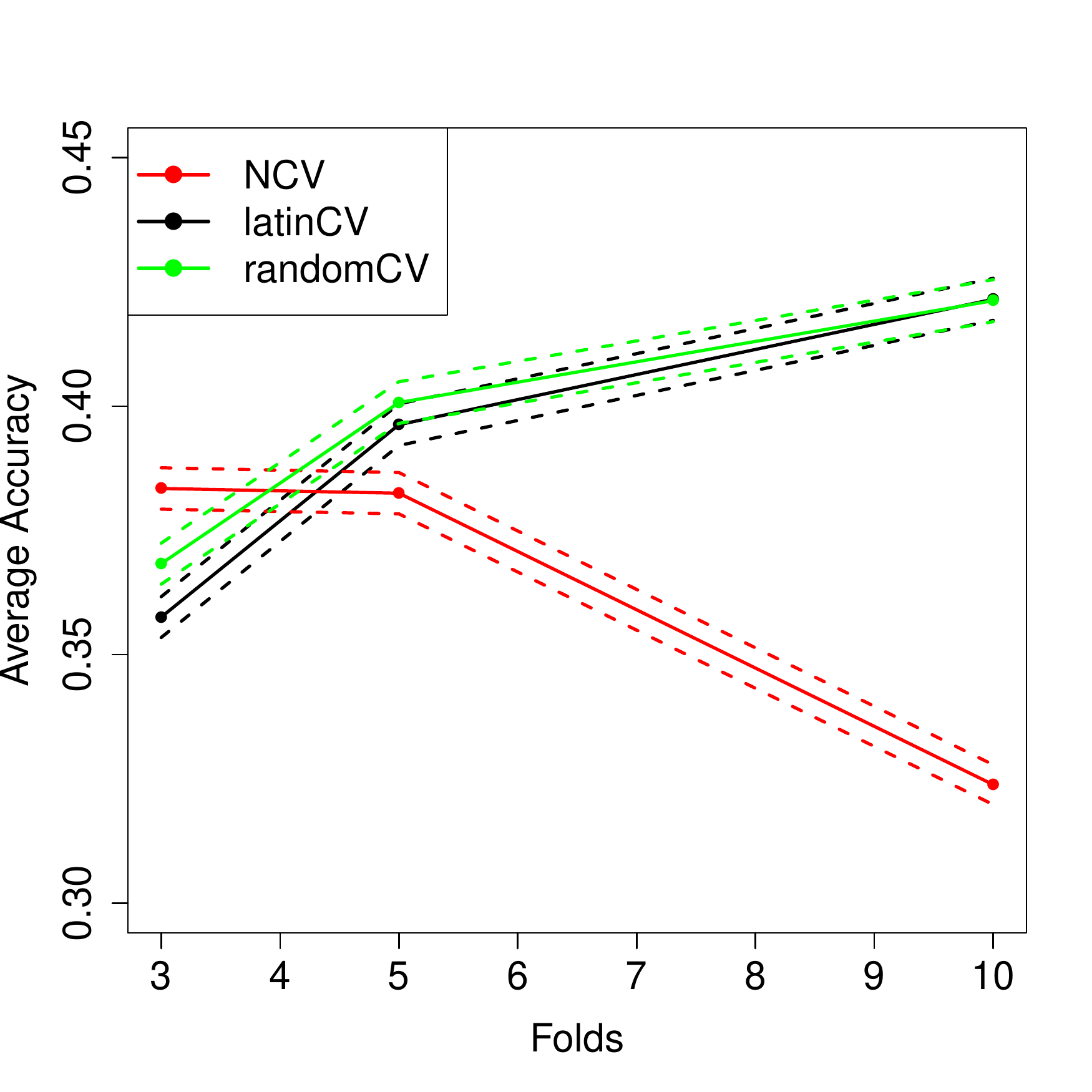}
\caption{Average Accuracy over all experimental cells.  Dashed lines represent a 95\% confidence interval.} \label{fold-accuracy-plot}
\end{center}
\end{figure}

We begin by comparing the performance of each cross-validation method as we vary the number of folds used for the cross-validation routines.  Figure \ref{fold-accuracy-plot} shows the average accuracy of the three methods as we vary the number of folds.  We see that for the NCV method we have reduced accuracy as we increase the number of folds, whereas the latinCV and randomCV methods increase in accuracy as we increase the number of folds.  This difference is because the NCV estimator increases in variance as we increase the number of folds, whereas the variance of latinCV and randomCV decreases as we increase the number of folds.  We discuss the variability of the cross-validation estimators in more detail in section \ref{subsec:cv-variance}.  For the remainder of this section we compare the 3 fold NCV method to the 10 fold latinCV and randomCV method so that we are comparing the best version of each cross-validation method.

In table \ref{cv-accuracy-nk} we consider the accuracy of the CV methods as we vary the number of nodes and blocks in the simulated network.  For this table, and all similar tables in this paper, we have averaged over all other experimental factors to obtain the averages presented in each cell.  We have highlighted the method that maximizes the model selection accuracy in each case.  NCV only has the largest model selection accuracy for the smallest networks, 30 nodes, with the two largest numbers of blocks, 4 and 5.  For both of those experimental conditions, the accuracy was very low for all model selection methods.  However, the latinCV and randomCV methods have little difference in accuracy across all cells of the table.  There also is no clear trend with latinCV and randomCV both having a larger accuracy in 7 cells each, and having the same accuracy in 4 cells.

Table \ref{cv-accuracy-rb} shows the accuracy of the cross-validation methods as we now vary the density and ratio parameters.  Again we see that latinCV and randomCV methods have indistinguishable performances.  However, averaging over all of the possible network and block sizes, NCV has a lower model recovery accuracy in all of the cells of this table.

% latex table generated in R 3.2.2 by xtable 1.8-2 package
% Sun May 01 12:38:16 2016
\begin{table}[ht]
\centering
\begin{tabular}{|rl|rrrrr|}
  \hline
\multicolumn{7}{|c|}{Cross-Validation Methods} \\ \hline
\multicolumn{2}{|c|}{Average Accuracy} & \multicolumn{5}{c|}{Blocks } \\ \hline
Nodes & Method & 1 & 2 & 3 & 4 & 5 \\ 
  \hline
   & network & 0.808 & 0.368 & 0.185 & \textbf{0.063} & \textbf{0.026} \\ 
    30 & random & \textbf{0.977} & 0.411 & 0.185 & 0.042 & 0.006 \\ 
     & latin & 0.975 & \textbf{0.426} & \textbf{0.190} & 0.048 & 0.007 \\ \hline
     
     & network & 0.941 & 0.411 & 0.237 & 0.129 & 0.066 \\ 
    60 & random & 0.997 & 0.444 & \textbf{0.266} & \textbf{0.142} & 0.067 \\ 
     & latin & \textbf{0.998} & \textbf{0.446} & 0.261 & 0.134 & \textbf{0.069} \\ \hline
     
    & network & 0.993 & 0.508 & 0.297 & 0.237 & 0.173 \\ 
   120 & random & \textbf{1.000} & \textbf{0.561} & 0.338 & \textbf{0.272} & \textbf{0.187} \\ 
    & latin & \textbf{1.000} & 0.558 & \textbf{0.343} & 0.271 & \textbf{0.187} \\ \hline
    
    & network & 0.932 & 0.723 & 0.431 & 0.323 & 0.324 \\ 
   300 & random & \textbf{1.000} & \textbf{0.786} & \textbf{0.498} & \textbf{0.415} & 0.346 \\ 
    & latin & \textbf{1.000} & \textbf{0.786} & 0.493 & 0.414 & \textbf{0.348} \\ 
   \hline
\end{tabular}
\caption{Comparing average accuracy across network and block sizes.  latinCV and randomCV have comparable accuracy in all cells of the table.  NCV only performs better in the 30 node networks with 4 or 5 blocks, where the average accuracy is very low for all methods.}
\label{cv-accuracy-nk}
\end{table}

% latex table generated in R 3.2.2 by xtable 1.8-2 package
% Sun May 01 12:39:16 2016
\begin{table}[ht]
\centering
\begin{tabular}{|rl|rrr|}
  \hline
\multicolumn{5}{|c|}{Cross-Validation Methods} \\ \hline
\multicolumn{2}{|c|}{Average Accuracy} & \multicolumn{3}{c|}{Ratio (r)} \\ \hline
Density (b) & Method & 3 & 4 & 5 \\ 
  \hline
         & NCV & 0.170 & 0.215 & 0.230 \\ 
0.010 & random & \textbf{0.174} & \textbf{0.232} & \textbf{0.252} \\ 
       & latin & 0.173 & 0.231 & \textbf{0.252} \\ \hline
 
         & NCV & 0.298 & 0.397 & 0.466 \\ 
0.050 & random & \textbf{0.315} & \textbf{0.423} & 0.503 \\ 
       & latin & 0.314 & 0.421 & \textbf{0.508} \\ \hline
 
         & NCV & 0.435 & 0.532 & 0.595 \\ 
0.100 & random & \textbf{0.469} & 0.593 & 0.705 \\ 
       & latin & 0.465 & \textbf{0.596} & \textbf{0.707} \\ 
   \hline
\end{tabular}
\caption{Comparing model selection accuracy as we vary the density and ratio parameters.  Again latinCV and randomCV have similar accuracy in all cells of the table, while NCV has lower accuracy in all cells of the table.}
\label{cv-accuracy-rb}
\end{table}

\subsubsection{Comparing Experimental Conditions}

\begin{figure}
\begin{center}
\includegraphics[width=3.0in]{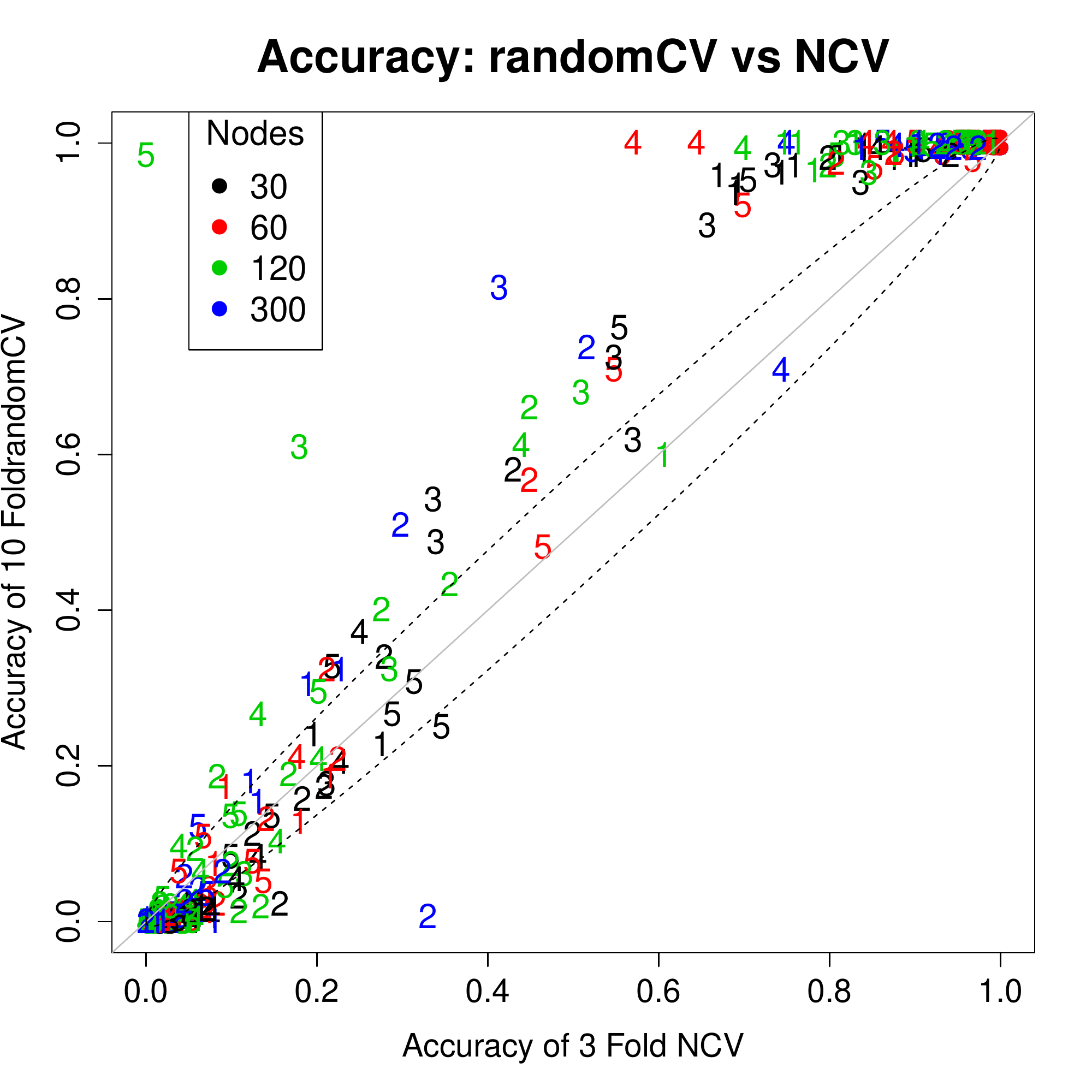}
\includegraphics[width=3.0in]{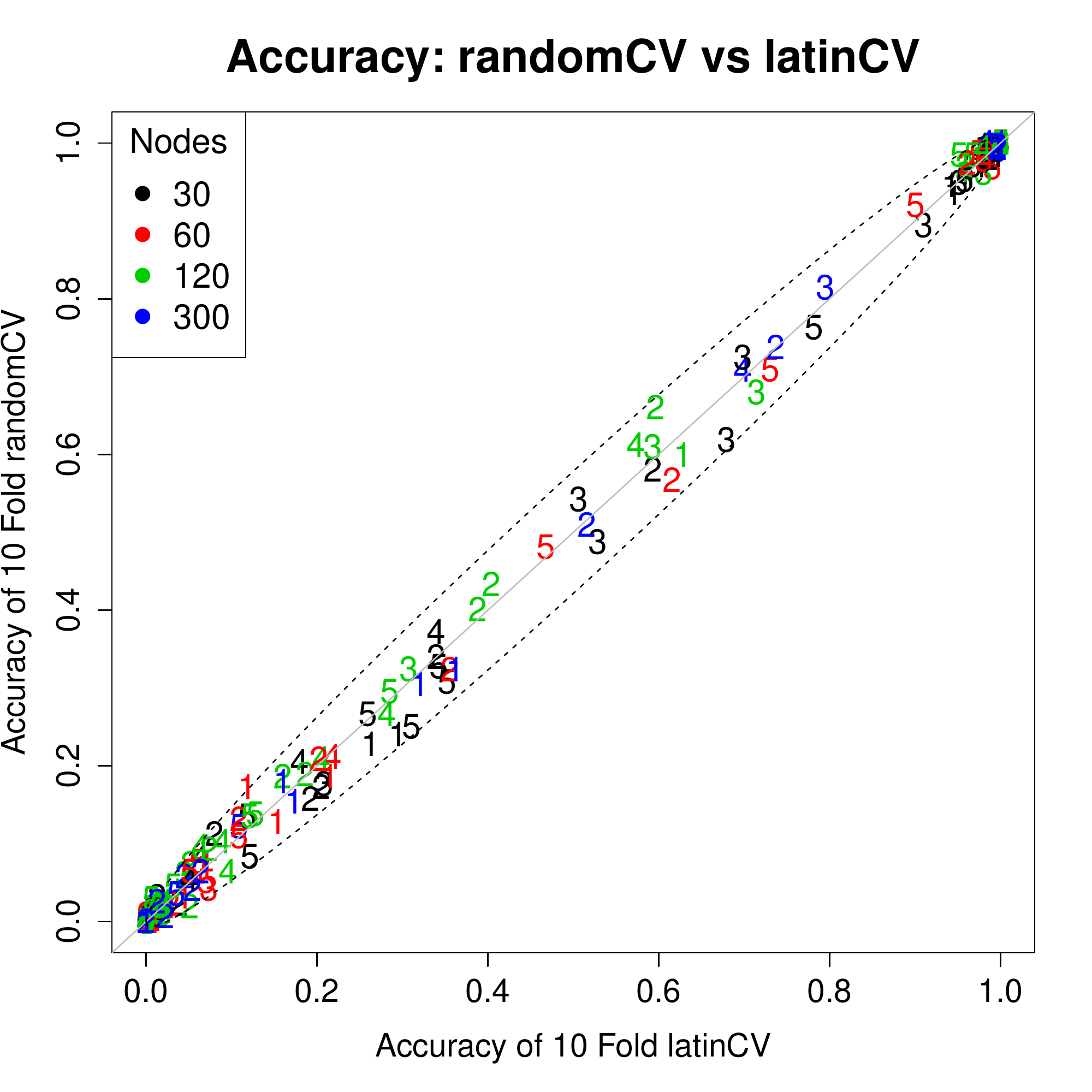}
\caption{Comparison of model selection accuracy for all experimental conditions.  Each point represents one of the block size schemes combined with the parameters $(n,k,b,r)$.  The number of nodes is represented by the color, and the number of blocks by the text.  The dashed line in both plots represents a 95\% confidence interval for the null hypothesis that the two accuracies are the same.  \textbf{Left:} Comparison of randomCV to NCV.  \textbf{Right:} Comparison of randomCV to latinCV.}
\label{cv-cells-plot}
\end{center}
\end{figure}

Figure \ref{cv-cells-plot} plots the model selection accuracy for the cross-validation methods against each other for each experimental condition.  Comparing randomCV to NCV, we can see that for many of the experimental conditions, randomCV has a significantly larger accuracy than NCV.  In general, when randomCV has an accuracy of 20\% or more, it tends to perform as well or better than NCV.  However, NCV does have a significantly larger accuracy in some cases where both methods have an accuracy below 20\%.  The reason for this discrepancy appears to be the increased variability of NCV, which allows it to capitalize on chance when the model selection problem is difficult, but causes it to be hurt by the same variability when the problem is simpler.  In contrast, there were no significant differences in accuracy when comparing latinCV and randomCV.

\subsubsection{MSE Comparison}

The last criterion we compare for the cross-validation methods is the mean-squared error (MSE) of the estimators chosen by each method.  Table \ref{cv-mse-table} shows how the MSE varies as we increase the number of nodes and blocks in the network.  As the network size increases, the MSE decreases significantly for all of the methods, while the MSE increases as more blocks are added to the network.  Again, latinCV and randomCV perform similarly at the MSE minimization task as well.  However, the estimators chosen by NCV have a larger MSE on average than the other two cross-validation methods for every cell in the table.  Thus, part of the reason that NCV has a lower accuracy than the other methods, is simply because it is doing a worse job of minimizing the mean-squared error that CV methods are designed to minimize.

% latex table generated in R 3.2.2 by xtable 1.8-2 package
% Sun May 01 12:45:24 2016
\begin{table}[ht]
\centering
\begin{tabular}{|rl|rrrrr|}
  \hline
\multicolumn{7}{|c|}{Cross-Validation Methods} \\ \hline
\multicolumn{2}{|c|}{Average MSE} & \multicolumn{5}{c|}{Blocks} \\ \hline
Nodes & Method & 1 & 2 & 3 & 4 & 5 \\ 
  \hline
       & NCV & 0.0014611 & 0.0061138 & 0.0084976 & 0.0092624 & 0.0090397 \\ 
        30 & random & \textbf{0.0002636} & 0.0048044 & \textbf{0.0069542} & \textbf{0.0080701} & 0.0079794 \\ 
         & latin & 0.0002739 & \textbf{0.0047723} & 0.0069571 & 0.0080712 & \textbf{0.0079173} \\ \hline
         
              & NCV & 0.0001226 & 0.0010749 & 0.0021496 & 0.0027772 & 0.0031560 \\ 
        60 & random & 0.0000427 & 0.0008281 & 0.0018486 & 0.0025117 & 0.0028966 \\ 
            & latin & 0.0000425 & 0.0008344 & 0.0018548 & 0.0025032 & 0.0029035 \\ \hline
         
        & NCV & 0.0000120 & 0.0002714 & 0.0006832 & 0.0010161 & 0.0013396 \\ 
 120 & random & \textbf{0.0000102} & \textbf{0.0002118} & \textbf{0.0006170} & \textbf{0.0009321} & 0.0012623 \\ 
      & latin & \textbf{0.0000102} & 0.0002140 & 0.0006188 & 0.0009359 & \textbf{0.0012536} \\ \hline

         & NCV & 0.0000334 & 0.0000563 & 0.0001712 & 0.0003100 & 0.0004128 \\ 
  300 & random & \textbf{0.0000016} & \textbf{0.0000479} & \textbf{0.0001600} & \textbf{0.0002917} & \textbf{0.0003942} \\ 
       & latin & \textbf{0.0000016} & 0.0000480 & 0.0001607 & 0.0002946 & 0.0003991 \\ 
   \hline
\end{tabular}
\caption{Comparing average MSE across network and block sizes.  NCV never minimizes the average MSE.}
\label{cv-mse-table}
\end{table}

For the rest of this section we compare latinCV to the alternative model selection methods.  We focus on latinCV because it performed better in accuracy than the other CV methods, but note that comparisons to randomCV would have similar results.

\subsection{Comparing latinCV and Information Criteria}

Now we compare the performance of the information criteria introduced in section \ref{ic-methods} to the 10 fold latinCV method.  Table \ref{ic-accuracy-nk} shows that AIC has a particularly low model selection accuracy for all experimental conditions, except for the smallest networks, 30 nodes, with the most blocks, 4 and 5.  Given that AIC has a model selection accuracy of 5\% or less in every other experimental condition, the version of AIC we have used here does not appear to be a viable option for performing accurate model selection.  BIC, however, is able to outperform latinCV for many of the experimental conditions.  %First we will discuss the strengths of BIC, and then we will discuss its weaknesses.

BIC has a remarkably high model selection accuracy for the networks with between 60 and 120 nodes, and between 2 and 4 blocks, sometimes getting 20\% more of the models correct than latinCV.  Similarly, for the networks with 300 nodes, and 3 or more blocks, BIC performs more accurately than latinCV.

However, BIC always performs worse when the true generative model had a single block, and thus no actual group structure.  Further, for single block models the model selection accuracy of BIC becomes worse, not better, as the size of the network grows.  You can also see that the accuracy never increases above 70\% for any cell, and for the 2 block models, the accuracy decreases from 120 node to 300 node networks.  latinCV, in contrast, always increases in accuracy as we increase the number of nodes, and hence the sample size used for model estimation.  In future experiments, it would be interesting to see if the accuracy for BIC begins decreasing for the 3, 4, and 5 block models as we increase the number of nodes in the network beyond 300 nodes.

Table \ref{ic-accuracy-rb} shows that BIC performs better than latinCV when the base tie density is small ($0.01$).  For both BIC and latinCV, model selection accuracy increases as the density and ratio parameters increase, but latinCV's accuracy increases more quickly, and thus latinCV is the preferred method for networks with larger base tie densities and a larger discrepancy between within and between block ties.
%
%
%
%\begin{itemize}
%\item AIC performs poorly overall, tending to prefer more complex models.
%\item BIC performs quite well overall
%\item BIC performed worse than CV for all Erdos Renyi Cases
%\item As $n$ increased, BIC also performed worse for 2 block models
%\end{itemize}
%
%\begin{itemize}
%\item BIC is the preferred method in 4 of the 9 cells
%\item randomCV performs better when the signal to noise ratio is larger (i.e. when r and b are larger)
%\end{itemize}

% latex table generated in R 3.2.2 by xtable 1.8-2 package
% Sun May 01 12:39:34 2016
\begin{table}[ht]
\centering
\small
\begin{tabular}{|rl|rrrrr|}
  \hline
\multicolumn{7}{|c|}{Information Criteria} \\ \hline
\multicolumn{2}{|c|}{Average Accuracy} & \multicolumn{5}{c|}{Blocks} \\ \hline
Nodes & Method & 1 & 2 & 3 & 4 & 5 \\ 
  \hline
     & latin & \textbf{0.975} & 0.426 & 0.190 & 0.048 & 0.007 \\
    30 & BIC & 0.834 & \textbf{0.629} & \textbf{0.261} & 0.008 & 0.000 \\ 
     & AIC & 0.007 & 0.001 & 0.026 & \textbf{0.175} & \textbf{0.427} \\  \hline
     & latin & \textbf{0.998} & 0.446 & 0.261 & 0.134 & \textbf{0.069} \\
    60 & BIC & 0.604 & \textbf{0.662} & \textbf{0.345} & \textbf{0.174} & 0.060 \\ 
     & AIC & 0.000 & 0.006 & 0.020 & 0.027 & 0.054 \\  \hline
    & latin & \textbf{1.000} & 0.558 & 0.343 & 0.271 & 0.187 \\
   120 & BIC & 0.349 & \textbf{0.702} & \textbf{0.506} & \textbf{0.330} & \textbf{0.245} \\ 
    & AIC & 0.000 & 0.000 & 0.000 & 0.002 & 0.002 \\ \hline
    & latin & \textbf{1.000} & \textbf{0.786} & 0.493 & 0.414 & 0.348 \\
   300 & BIC & 0.158 & 0.399 & \textbf{0.532} & \textbf{0.493} & \textbf{0.451} \\ 
    & AIC & 0.000 & 0.000 & 0.000 & 0.000 & 0.001 \\ 
   \hline
\end{tabular}
\caption{Comparing average accuracy across network and block sizes.  BIC has the largest accuracy in many of the cells, but performs poorly for one block networks and large 2 block networks.  Accuracy decreases as the network grows for one and two block models for BIC.  AIC has less than 5\% accuracy for all but 2 cells.}
\label{ic-accuracy-nk}
\end{table}

% latex table generated in R 3.2.2 by xtable 1.8-2 package
% Sun May 01 12:39:43 2016
\begin{table}[ht]
\centering
\small
\begin{tabular}{|rl|rrr|}
  \hline
\multicolumn{5}{|c|}{Information Criteria} \\ \hline
\multicolumn{2}{|c|}{Average Accuracy} & \multicolumn{3}{c|}{Ratio (r)} \\ \hline
Density (b) & Method & 3 & 4 & 5 \\ 
  \hline
 & latin & 0.173 & 0.231 & 0.252 \\
  0.010 & BIC & \textbf{0.208} & \textbf{0.234} & \textbf{0.262} \\ 
 & AIC & 0.032 & 0.019 & 0.012 \\ \hline
& latin & \textbf{0.314} & \textbf{0.421} & \textbf{0.508} \\
  0.050 & BIC & 0.291 & 0.376 & 0.456 \\ 
 & AIC & 0.033 & 0.033 & 0.037 \\ \hline
 & latin & \textbf{0.465} & \textbf{0.596} & \textbf{0.707} \\ 
  0.100 & BIC & 0.412 & 0.503 & 0.591 \\ 
 & AIC & 0.031 & 0.035 & 0.042 \\ 
   \hline
\end{tabular}
\caption{Comparing model selection accuracy as we vary density and ratio parameters.  BIC performs well for networks with a low density of ties, but latinCV maximizes model recovery accuracy for networks with a higher tie density.  AIC has low accuracy in all cases.}
\label{ic-accuracy-rb}
\end{table}

\subsubsection{MSE Comparison}

Table \ref{ic-mse-table} shows that BIC is actually able to do a better job minimizing MSE in many of the cases where BIC also had a larger model selection accuracy.  This result is particularly interesting given that latinCV is attempting to minimize an estimate of MSE directly.  The results for MSE very closely mirror those for model selection accuracy, with latinCV always performing better for 1 block models, and latinCV performing better for 2 block models when there are enough nodes in the network.

% latex table generated in R 3.2.2 by xtable 1.8-2 package
% Sun May 01 12:46:23 2016
\begin{table}[ht]
\centering
\begin{tabular}{|rl|rrrrr|}
  \hline
\multicolumn{7}{|c|}{Information Criteria} \\ \hline
\multicolumn{2}{|c|}{Average MSE} & \multicolumn{5}{c|}{Blocks} \\ \hline
Nodes & Method & 1 & 2 & 3 & 4 & 5 \\ 
  \hline
         & latin & \textbf{0.0002739} & 0.0047723 & 0.0069571 & 0.0080712 & 0.0079173 \\
        30 & BIC & 0.0013484 & \textbf{0.0042020} & \textbf{0.0064035} & \textbf{0.0078918} & \textbf{0.0078397} \\ 
         & AIC & 0.0208539 & 0.0176481 & 0.0169309 & 0.0172273 & 0.0174522 \\ \hline
         
          & latin & \textbf{0.0000425} & 0.0008344 & 0.0018548 & 0.0025032 & 0.0029035 \\
        60 & BIC & 0.0007910 & \textbf{0.0007939} & \textbf{0.0016801} & \textbf{0.0024156} & 0.0029063 \\ 
         & AIC & 0.0130959 & 0.0080541 & 0.0068322 & 0.0067750 & 0.0068175 \\ \hline
         
          & latin & \textbf{0.0000102} &\textbf{0.0002140} & 0.0006188 & 0.0009359 & 0.0012536 \\
       120 & BIC & 0.0005094 & 0.0003081 & \textbf{0.0006115} & \textbf{0.0009116} & \textbf{0.0011995} \\ 
        & AIC & 0.0061639 & 0.0040539 & 0.0034158 & 0.0031755 & 0.0031596 \\ \hline
        
       & latin & \textbf{0.0000016} & \textbf{0.0000480} & 0.0001607 & 0.0002946 & 0.0003991 \\ 
       300 & BIC & 0.0002914 & 0.0001406 & \textbf{0.0001553} & \textbf{0.0002647} & \textbf{0.0003544} \\ 
        & AIC & 0.0007387 & 0.0006394 & 0.0005968 & 0.0006170 & 0.0006413 \\ 
   \hline
\end{tabular}
\caption{Comparing average MSE across network and block sizes.  BIC is able to minimize MSE on average when there are many blocks and few nodes.  MSE is minimized by latinCV for one block models and two block models with 120 or more nodes.}
\label{ic-mse-table}
\end{table}

\subsection{Comparing latinCV and Community Detection}

The last two model selection methods we consider are modularity maximization and infomap.  Table \ref{cd-accuracy-nk} compares the model selection accuracy of these two methods to that of latinCV.  The infomap method never performs better than both modularity and latinCV, though it does have a larger accuracy than latinCV for 30 node networks with 4 or 5 blocks.  The modularity maximization routine has characteristics similar to BIC.  The modularity method performs moderately well for smaller networks with a larger number of blocks.  However, the modularity has even greater accuracy for the 4 and 5 block models with less than 300 nodes than BIC.  Unfortunately, the modularity criterion never selected a clustering with a single community in all of the experiments in our simulation study.  In fact, a single block model with a tie probability of greater than 90\% was needed for the modularity criterion to be minimized by a single community.  The modularity criterion also had decreasing accuracy as the size of the network was increased in some cases, and never had an accuracy greater than 50\% for networks with more than 30 nodes.

Considering table \ref{cd-accuracy-rb} we can see that both community detection methods increased in accuracy as the density and ratio parameters increased.  However, for each combination of parameters, averaging over the number of nodes and blocks in the network, latinCV was the preferred method.  This results is in part due to the fact that the modularity routine had $0\%$ accuracy for all examples generated from single block models.

% latex table generated in R 3.2.2 by xtable 1.8-2 package
% Sun May 01 12:40:02 2016
\begin{table}[ht]
\centering
\begin{tabular}{|rl|rrrrr|}
  \hline
\multicolumn{7}{|c|}{Community Detection Methods} \\ \hline
\multicolumn{2}{|c|}{Average Accuracy} & \multicolumn{5}{c|}{Blocks} \\ \hline
Nodes & Method & 1 & 2 & 3 & 4 & 5 \\ 
  \hline
     & latin & \textbf{0.975} & \textbf{0.426} & 0.190 & 0.048 & 0.007 \\
    30 & modularity & 0.000 & 0.227 & \textbf{0.259} & \textbf{0.639} & \textbf{0.320} \\ 
     & infomap & 0.671 & 0.104 & 0.149 & 0.116 & 0.095 \\ \hline
     
       & latin &\textbf{ 0.998} & \textbf{0.446} & \textbf{0.261} & 0.134 & 0.069 \\
    60 & modularity & 0.000 & 0.260 & 0.244 & \textbf{0.415} & \textbf{0.284} \\ 
     & infomap & 0.677 & 0.020 & 0.118 & 0.075 & 0.038 \\ \hline
     
     & latin & \textbf{1.000} & \textbf{0.558} & 0.343 & 0.271 & 0.187 \\
   120 & modularity & 0.000 & 0.344 & \textbf{0.381} & \textbf{0.434} & \textbf{0.289} \\ 
    & infomap & 0.837 & 0.004 & 0.175 & 0.124 & 0.071 \\ \hline
    
    & latin & \textbf{1.000} & \textbf{0.786} & \textbf{0.493} & \textbf{0.414} & 0.348 \\ 
   300 & modularity & 0.000 & 0.444 & 0.459 & 0.405 & \textbf{0.353} \\ 
    & infomap & \textbf{1.000} & 0.031 & 0.167 & 0.137 & 0.112 \\ 
   \hline
\end{tabular}
\caption{Comparing model selection accuracy across network and block size.  The modularity maximization method has larger accuracy than latinCV for networks with many blocks and few enough nodes.  Modularity always finds a community assignment with more than one block.  Infomap has lower accuracy than latinCV in all but two cells, corresponding to the smallest networks with the most blocks (4 and 5).}
\label{cd-accuracy-nk}
\end{table}

%\begin{itemize}
%\item Averaging over number of nodes and blocks, latinCV is preferred for all selections of $r$ and $b$.
%\end{itemize}

% latex table generated in R 3.2.2 by xtable 1.8-2 package
% Sun May 01 12:40:11 2016
\begin{table}[ht]
\centering
\begin{tabular}{|rl|rrr|}
  \hline
  \multicolumn{5}{|c|}{Community Detection Methods} \\ \hline
\multicolumn{2}{|c|}{Average Accuracy} & \multicolumn{3}{c|}{Ratio (r)} \\ \hline
Density (b) & Method & 3 & 4 & 5 \\ 
  \hline
 & latin & \textbf{0.173} & \textbf{0.231 }&\textbf{ 0.252} \\
  0.010 & modularity & 0.000 & 0.010 & 0.079 \\ 
 & infomap & 0.084 & 0.105 & 0.121 \\ \hline
 & latin & \textbf{0.314} & \textbf{0.421} &\textbf{ 0.508} \\
  0.050 & modularity & 0.283 & 0.384 & 0.439 \\ 
 & infomap & 0.172 & 0.225 & 0.359 \\ \hline
 & latin & \textbf{0.465} & \textbf{0.596 }& \textbf{0.707} \\ 
  0.100 & modularity & 0.406 & 0.480 & 0.506 \\ 
 & infomap & 0.137 & 0.202 & 0.357 \\ 
   \hline
\end{tabular}
\caption{Comparing model selection accuracy as we vary density and ratio parameters.  latinCV is the most accurate method across all cells.  This result is in part due to the modularity method's inability to correctly select single block models.}
\label{cd-accuracy-rb}
\end{table}

\subsubsection{MSE Comparison}

Table \ref{cd-mse-table} shows the mean-squared error for the community detection methods. Unlike BIC, the community detection methods were never able to minimize the mean-squared error more effectively than latinCV.  Even in the cases where the modularity maximization method was able to more accurately recover the number of blocks, it was at the cost of a significant increase in MSE.  Both community detection methods always had at least a 30\% increase in MSE, with more than triple the mean-squared error in many cases.

% latex table generated in R 3.2.2 by xtable 1.8-2 package
% Sun May 01 12:47:50 2016
\begin{table}[ht]
\centering
\begin{tabular}{|rl|rrrrr|}
  \hline
\multicolumn{7}{|c|}{Community Detection Methods} \\ \hline
\multicolumn{2}{|c|}{Average MSE} & \multicolumn{5}{c|}{Blocks} \\ \hline
Nodes & Method & 1 & 2 & 3 & 4 & 5 \\ 
  \hline
           & latin & \textbf{0.0002739} & \textbf{0.0047723} & \textbf{0.0069571} & \textbf{0.0080712} & \textbf{0.0079173} \\
        30 & modularity & 0.0106576 & 0.0087314 & 0.0114866 & 0.0114285 & 0.0123177 \\ 
         & infomap & 0.0016157 & 0.0138215 & 0.0128692 & 0.0127446 & 0.0122040 \\ \hline
         
              & latin & \textbf{0.0000425} & \textbf{0.0008344} & \textbf{0.0018548} & \textbf{0.0025032} & \textbf{0.0029035} \\
        60 & modularity & 0.0057105 & 0.0029073 & 0.0034274 & 0.0036905 & 0.0042902 \\ 
         & infomap & 0.0009732 & 0.0104082 & 0.0076307 & 0.0077168 & 0.0078128 \\ \hline
         
        & latin & \textbf{0.0000102} & \textbf{0.0002140} & \textbf{0.0006188} & \textbf{0.0009359} & \textbf{0.0012536} \\
       120 & modularity & 0.0028109 & 0.0009979 & 0.0014216 & 0.0015433 & 0.0018438 \\ 
        & infomap & 0.0002337 & 0.0096175 & 0.0066107 & 0.0068598 & 0.0068248 \\ \hline
        
         & latin & \textbf{0.0000016} & \textbf{0.0000480} & \textbf{0.0001607} & \textbf{0.0002946} & \textbf{0.0003991} \\ 
       300 & modularity & 0.0010832 & 0.0002632 & 0.0006360 & 0.0008597 & 0.0008760 \\ 
        & infomap & 0.0000016 & 0.0097989 & 0.0072673 & 0.0071816 & 0.0065234 \\ 
   \hline
\end{tabular}
\caption{Comparing MSE as we vary network and block size.  We see that the community detection methods do a poor job selecting models that minimize MSE for all combinations of network and block size.}
\label{cd-mse-table}
\end{table}

\subsection{Confusion Matrices}

Now we consider what alternative models are being chosen by each of the methods.  From the previous tables and figures we have been able to get a sense for how accurate each method is, but we do not know if the methods tend to be overfitting, choosing a larger number of blocks, or underfitting, choosing a smaller number of blocks.  %restrict our attention to the best method in each category - latinCV, BIC, and modularity.

Table \ref{latin-network-confusion} shows the confusion matrices for latinCV and NCV.  We can see that both methods frequently underfit.  For networks with 3 or more blocks, both CV methods selected a model with a single block more often than the true number of blocks, or any other number of blocks.  The difference between the two methods is that the NCV has greater variability in models selected.  NCV more often chooses larger models than the truth, and less often chooses the correct number of blocks.

Table \ref{bic-mod-conf} compares the confusion matrices of the alternative model selection procedures BIC and modularity.  As we saw in the previous sections, BIC overfits for the one block models, but now we can say that it also tends to overfit for the 2 block models as well.  For networks with 3 or more blocks, BIC underfits more than it overfits, and tends to prefer 2 block models to the single block model.  Modularity, by comparison, has a greater tendency to overfit, detecting more blocks than the truth more than 40\% of the time.  Modularity also selected some models with more than 11 blocks, that are not reported in this table.

% latex table generated in R 3.2.2 by xtable 1.8-2 package
% Sun May 01 13:27:00 2016
\begin{table}[ht]
\centering
\small
\begin{tabular}{|r|rrrrr|}
  \hline
  \multicolumn{6}{|c|}{latinCV Confusion Matrix} \\ \hline
 & \multicolumn{5}{|c|}{True Blocks} \\
$\hat{K}$ & 1 & 2 & 3 & 4 & 5 \\ 
  \hline
   1 & \textbf{0.995} & 0.315 & 0.354 & 0.418 & 0.457 \\ 
     2 & 0.004 & \textbf{0.565} & 0.166 & 0.162 & 0.160 \\ 
     3 & 0.001 & 0.065 & \textbf{0.334} & 0.069 & 0.060 \\ 
     4 & 0.000 & 0.027 & 0.056 & \textbf{0.232} & 0.040 \\ 
     5 & 0.000 & 0.012 & 0.038 & 0.048 & \textbf{0.166} \\ 
     6 & 0.000 & 0.008 & 0.021 & 0.024 & 0.040 \\ 
     7 & 0.000 & 0.004 & 0.011 & 0.014 & 0.027 \\ 
     8 & 0.000 & 0.003 & 0.007 & 0.010 & 0.016 \\ 
     9 & 0.000 & 0.001 & 0.005 & 0.009 & 0.015 \\ 
    10 & 0.000 & 0.000 & 0.004 & 0.008 & 0.013 \\ 
    11 & 0.000 & 0.000 & 0.003 & 0.005 & 0.006 \\ 
   \hline
\end{tabular}
\quad
% latex table generated in R 3.2.2 by xtable 1.8-2 package
% Sun May 01 13:27:01 2016
%\begin{table}[ht]
%\centering
\begin{tabular}{|r|rrrrr|}
  \hline
  \multicolumn{6}{|c|}{NCV Confusion Matrix} \\ \hline
 & \multicolumn{5}{|c|}{True Blocks} \\
$\hat{K}$ & 1 & 2 & 3 & 4 & 5 \\
  \hline
   1 & \textbf{0.921} & 0.281 & 0.320 & 0.376 & 0.412 \\ 
     2 & 0.035 & \textbf{0.514} & 0.157 & 0.155 & 0.151 \\ 
     3 & 0.017 & 0.103 & \textbf{0.297} & 0.081 & 0.079 \\ 
     4 & 0.009 & 0.048 & 0.088 & \textbf{0.200} & 0.049 \\ 
     5 & 0.006 & 0.020 & 0.049 & 0.079 & \textbf{0.158} \\ 
     6 & 0.004 & 0.011 & 0.030 & 0.038 & 0.054 \\ 
     7 & 0.003 & 0.007 & 0.018 & 0.025 & 0.037 \\ 
     8 & 0.001 & 0.005 & 0.013 & 0.017 & 0.025 \\ 
     9 & 0.002 & 0.005 & 0.011 & 0.012 & 0.017 \\ 
    10 & 0.001 & 0.003 & 0.009 & 0.011 & 0.011 \\ 
    11 & 0.001 & 0.003 & 0.008 & 0.007 & 0.005 \\ 
   \hline
\end{tabular}
\caption{Confusion matrices for latinCV and NCV.  We show the percentage of all assignments for each true number of blocks.  The columns represent the true generative model, and the rows the model chosen by model selection method.  Overall NCV selects a wider variety of models.}
\label{latin-network-confusion}
\end{table}

%% latex table generated in R 3.2.2 by xtable 1.8-2 package
%% Sun May 01 13:27:00 2016
\begin{table}[ht]
\centering
\small
\begin{tabular}{|r|rrrrr|}
  \hline
  \multicolumn{6}{|c|}{BIC Confusion Matrix} \\ \hline
 & \multicolumn{5}{|c|}{True Blocks} \\
$\hat{K}$ & 1 & 2 & 3 & 4 & 5 \\ 
  \hline
   1 &\textbf{0.421} & 0.100 & 0.121 & 0.154 & 0.183 \\ 
     2 & 0.410 & \textbf{0.593} & 0.308 & 0.335 & 0.353 \\ 
     3 & 0.083 & 0.201 & \textbf{0.425} & 0.145 & 0.128 \\ 
     4 & 0.018 & 0.051 & 0.077 & \textbf{0.273} & 0.075 \\ 
     5 & 0.067 & 0.037 & 0.044 & 0.050 & \textbf{0.206} \\ 
     6 & 0.001 & 0.014 & 0.015 & 0.023 & 0.028 \\ 
     7 & 0.000 & 0.003 & 0.004 & 0.010 & 0.014 \\ 
     8 & 0.000 & 0.000 & 0.003 & 0.006 & 0.010 \\ 
     9 & 0.000 & 0.000 & 0.002 & 0.003 & 0.002 \\ 
    10 & 0.000 & 0.000 & 0.000 & 0.000 & 0.000 \\ 
    11 & 0.000 & 0.000 & 0.000 & 0.000 & 0.000 \\ 
   \hline
\end{tabular}
%\end{table}
\quad
%
%% latex table generated in R 3.2.2 by xtable 1.8-2 package
%% Sun May 01 13:27:01 2016
%\begin{table}[ht]
%\centering
\begin{tabular}{|r|rrrrr|}
  \hline
  \multicolumn{6}{|c|}{Modularity Confusion Matrix} \\ \hline
 & \multicolumn{5}{|c|}{True Blocks} \\
$\hat{K}$ & 1 & 2 & 3 & 4 & 5 \\ 

  \hline
		1 & \textbf{0.000} & 0.000 & 0.000 & 0.000 & 0.000 \\
   2 & 0.002 & \textbf{0.327} & 0.053 & 0.077 & 0.071 \\ 
     3 & 0.052 & 0.150 & \textbf{0.343} & 0.095 & 0.092 \\ 
     4 & 0.628 & 0.246 & 0.286 & \textbf{0.458} & 0.216 \\ 
     5 & 0.145 & 0.079 & 0.083 & 0.111 & \textbf{0.311} \\ 
     6 & 0.108 & 0.060 & 0.049 & 0.049 & 0.074 \\ 
     7 & 0.051 & 0.070 & 0.071 & 0.058 & 0.060 \\ 
     8 & 0.012 & 0.048 & 0.069 & 0.082 & 0.081 \\ 
     9 & 0.001 & 0.016 & 0.033 & 0.045 & 0.057 \\ 
    10 & 0.000 & 0.003 & 0.009 & 0.017 & 0.026 \\ 
    11 & 0.000 & 0.000 & 0.003 & 0.005 & 0.008 \\ 
   \hline
\end{tabular}
\caption{Confusion matrices for BIC and modularity.  We show the percentage of all assignments for each true number of blocks.  The columns represent the true generative model, and the rows the model chosen by model selection method.  BIC tends to overfit when the true number of blocks is one or two, but underfits when there are three or more blocks.  Modularity has a tendency to overfit for all true number of blocks.}
\label{bic-mod-conf}
\end{table}

\clearpage

\section{Discussion}
\label{sec:discussion}

Over all of the experimental conditions, the cross-validation methods performed better than the alternative methods.  In particular, latinCV and randomCV were more accurate than the NCV method introduced previously.  The modularity maximization method showed remarkably large model recovery accuracy for some of the more difficult model selection problems, but had a tendency to overfit in general.  BIC also had difficulty with overfitting for the 1 and 2 block models, which worsened as we increased the number of nodes in the network.  In this section we discuss some underlying causes for the results we observed.

\subsection{Variance of Cross-validation Estimators}
\label{subsec:cv-variance}

A key result that we found was that the NCV method had an overall lower model selection accuracy than latinCV and randomCV.  The primary difference between these cross-validation methods is the amount of data used to evaluate each model.  In particular, for $V$-fold cross-validation, NCV assigns less than a $\frac{1}{V}$ fraction of the edge indicators to validation sets (folds that are used to evaluate the model).  Since our cross-validation risk estimators are an average over the loss on the validation sets, we would expect the variance of the CV risk estimator to increase as we decrease the size of the validation sets.  

\begin{figure}
\begin{center}
\includegraphics[width=5.5in]{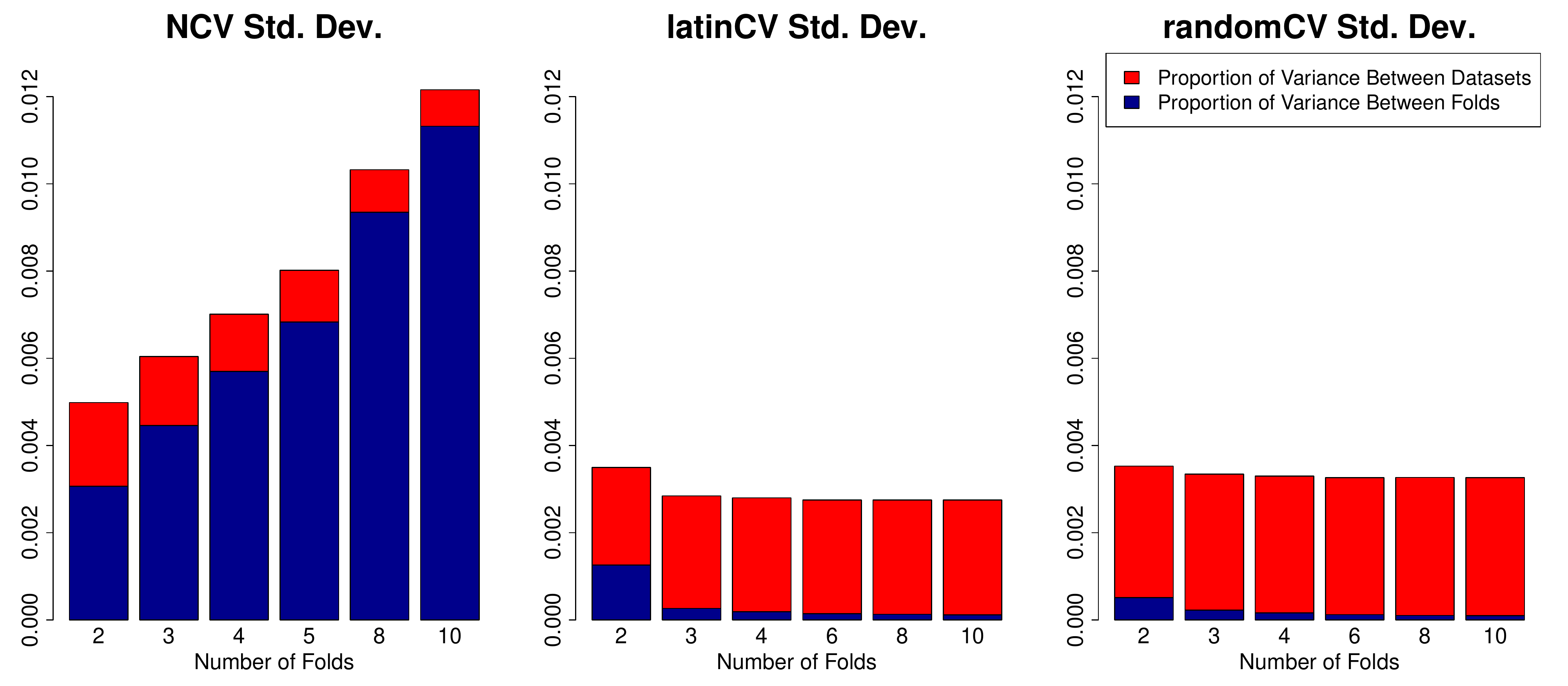}
\caption{Standard deviation of cross-validation risk estimates for NCV, latinCV, and randomCV.  We used CV to estimate the risk of using a 3 block model to estimate tie probabilities for a network with 60 nodes and two blocks.  The blue regions is the proportion of the overall variance due to varying the fold assignment, while the red regions represent the proportion of variance due to varying the network.}
\label{var-comp}
\end{center}
\end{figure}

Figure \ref{var-comp} shows the variance of the three cross-validation estimators of the risk for using a 3 block SBM to fit the data.  We simulated 100 networks from a 60 node network with 2 blocks, and used 100 different fold assignments for each network.  This simulation allows us to compare the variability of the estimates across different networks, and across different fold assignments, to assess the overall variation as well as the contributors to that variation.

Figure \ref{var-comp} shows clearly that the standard deviation of the NCV risk estimate increases as we increase the number of folds.  We have done similar simulations for many choices of the parameters considered in the simulation study, and have found the same result every time for NCV.  We have also colored the barplots based on the contribution of the varying fold assignments and the varying networks to the overall variance of the estimators.  With this separation, it is clear that most of the variability of the NCV estimator is due to the fold assignment procedure, and not the data variation.

For both latinCV and randomCV, we can see that the standard deviation decreases as we increase the number of folds.  In this example, latinCV has smaller standard deviation than randomCV, but similar to the results in the larger simulation study, there are cases where randomCV has a smaller standard deviation as well.  However, we would expect latinCV to have a smaller variance in general, due to the balancing of folds across nodes latinCV implements.  In future work we would like to consider more extensively the explicit bias and variance of these methods, both analytically and empirically.

\begin{figure}
\begin{center}
\includegraphics[width=3.5in]{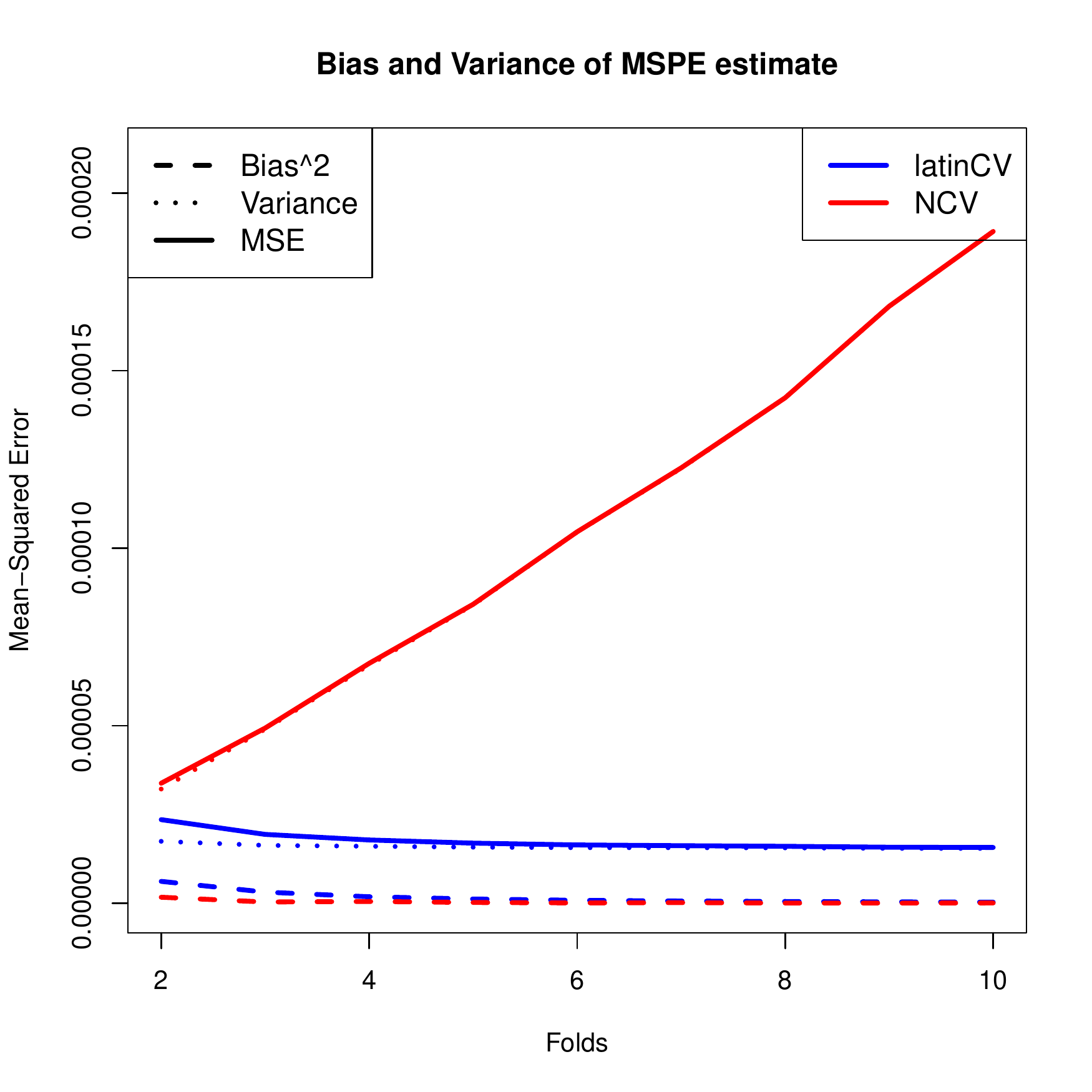}
\caption{The solid lines represent the mean squared error of latinCV (blue) and NCV (red) when estimating the true risk for using a three block model to estimate tie probabilities for a network 60 nodes and two blocks.  The wide dashed lines represent the squared bias, and the dotted lines represent the variance for the two cross-validation methods.}
\label{bias-var-cv}
\end{center}
\end{figure}

We also considered the bias of the cross-validation estimators as we varied the number of blocks.  Figure \ref{bias-var-cv} shows the variance and squared bias of the latinCV and NCV risk estimates as we increase the number of folds.  When added together, these two quantities equal the mean-squared error between the risk estimates and the true risk.  From this plot we can see that latinCV has a larger bias for 5 or fewer folds, but quickly reduces to near zero as we increase the number of folds.  Thus, for a sufficient number of folds, the bias is negligible when compared to the variance of the estimators.  This means the primary difference between the methods is the variability, which latinCV and randomCV are able to better minimize.  The fact that NCV is a more variable estimator of risk also explains the fact that NCV, selected a wider variety of models than latinCV in the simulation study.

%\begin{itemize}
%\item What's different about these estimators?
%\item The primary difference is variance.
%\item As we increase the number of folds, NCV has greatly increased variance, but variance decreases for randomCV
%\item The other thing that could affect model selection is the bias, but the bias goes to zero relatively quickly, and is much smaller than the variance, so the variance is the primary concern.
%\end{itemize}

\subsection{Using BIC for Network Models}

When we introduced our definitions for BIC and AIC, we noted that the choice for the degrees of freedom was not clear for latent variable models.  While the version of BIC we chose did perform better than the other versions mentioned in section \ref{sec:methods}, it is still possible that the degrees of freedom or sample size is misspecified.  The most troubling result with BIC was its inability to accurately select a model with no block structure.  It was also concerning that as the number of nodes increased, BIC performed worse for networks with 2 communities.  For 300 node networks with 2 blocks, BIC never selected a 1 block model, and selected models with more than 2 blocks 60\% of the time, suggesting that the BIC is underpenalizing complex models as the number of nodes in the network grows.

We did notice, however, that BIC was able to choose models with a lower average MSE than the latinCV method in some cases.  This result is particularly surprising because latinCV explicitly minimizes an estimate of the MSE.  One possible explanation for this difference could be the variability of the latinCV estimate of risk.  As we increase the number of blocks in the estimating model, the variability of the latinCV estimator of risk increases.  This increase in variance can cause latinCV to more often prefer simpler models that in truth have a larger mean-squared error.  Overall, BIC performed comparably to the latinCV estimator when the network truly had more than a single block, but the performance as the number of nodes increases suggests that this specification for BIC may be inaccurate for large networks.

\subsection{Modularity Maximization}

The modularity maximization method was able to perform remarkably well when the number of nodes per block was low.  However, in many of these cases, the minimizer of the mean-squared error turns out to not be the true model.  In particular, table \ref{true-risk-min} shows the actual risk minimizers for the 30 and 60 node networks with even sized blocks.  For these networks, modularity had better accuracy than latinCV for the 4 and 5 block models.  However, for the 30 node networks, selecting the true risk minimizer would never have resulted in a choice of the 5 block model, and only once would it have correctly resulted in the choice of the 4 block model.  Similarly for 60 node networks, the risk minimizer was only the true model twice for 5 block models, and 3 times for the 4 block models.

We saw in our results section that the modularity method was not selecting models that minimized mean-squared error.  Thus, the fact that it was selecting models that were correct, but did not actually minimize the MSE is not surprising.  However, this result suggests that we should not use modularity maximization methods if our goal is to select models with minimal MSE.  The modularity criterion had a tendency to select more complex models in general. When the true model was complex, but the risk minimizer was simple, this resulted in modularity performing better than expected, but when the true model was simple, the result was modularity performing poorly.

% latex table generated in R 3.2.2 by xtable 1.8-2 package
% Mon May 02 12:17:41 2016
\begin{table}[ht]
\centering
\begin{tabular}{|l|rrrrr|}
  \hline
   \multicolumn{6}{|c|}{30 Node Networks} \\ \hline
  & \multicolumn{5}{|c|}{True Blocks} \\ \hline
Risk Min. & 1 & 2 & 3 & 4 & 5 \\ 
  \hline
1 &   9 &   4 &   7 &   8 &   8 \\ 
  2 &   0 &   5 &   0 &   0 &   0 \\ 
  3 &   0 &   0 &   2 &   0 &   0 \\ 
  4 &   0 &   0 &   0 &   1 &   1 \\ 
  5 &   0 &   0 &   0 &   0 &   0 \\ 
   \hline
\end{tabular}
\begin{tabular}{|l|rrrrr|}
  \hline
   \multicolumn{6}{|c|}{60 Node Networks} \\ \hline
  & \multicolumn{5}{|c|}{True Blocks} \\ \hline
Risk Min. & 1 & 2 & 3 & 4 & 5 \\ 
  \hline
1 &   9 &   3 &   4 &   6 &   7 \\ 
  2 &   0 &   6 &   0 &   0 &   0 \\ 
  3 &   0 &   0 &   5 &   0 &   0 \\ 
  4 &   0 &   0 &   0 &   3 &   0 \\ 
  5 &   0 &   0 &   0 &   0 &   2 \\ 
   \hline
\end{tabular}
\caption{Confusion matrices for the true number of blocks vs. the number of blocks for the model that minimized the true risk.  \textbf{Left:} Confusion matrix for 30 nodes.  \textbf{Right:} Confusion matrix for 60 nodes.}
\label{true-risk-min}
\end{table}

\clearpage

\section{Conclusions}

In this paper we examined various cross-validation methods for networks, and compared them to alternative model selection procedures for networks.  We found that the specific fold assignment scheme, including the number of folds, is important for both model selection accuracy and risk estimation.  In particular we found that cross-validation methods that use all of the edge indicators for validation, including latinCV and randomCV, are preferable to those that do not, such as NCV.

We also found that using at least 10 folds increases the model selection accuracy for both latinCV and randomCV.  However, when we are using 10 folds to perform model selection, we found little difference between latinCV and randomCV in terms of accuracy and risk estimation.  This result suggests that the balance of fold assignments across the adjacency matrix is not important for accurate risk estimation, provided there are a sufficient number of folds.

Comparing the best of the cross-validation methods to alternative model selection procedures, we found that both the Akaike Informaiton Criterion and the infomap method were inaccurate at recovering the true number of blocks.  The Bayesian Informtion Criterion (BIC) performed accurately in some cases, but had issues with overfitting which seemed to be exacerbated as the size of the networks grew.  The common community detection method of modularity maximization also had issues with overfitting, choosing models with 2 or more additional blocks more than 25\% of the time.

Ultimately, we believe that cross-validation is the best model selection method available right now for network data.  We have shown that it performs more accurate model selection than alternative methods over a range of generative parameters for the Stochastic Block Model.  Further, the cross-validation methods defined here can also be used for any model with conditionally independent ties, extending the method to more complex models such as the Latent Space Model, the Mixed Membership Stochastic Block Model and the Degree-Corrected Stochastic Block Model, as well as any model involving covariates.  This flexibility, combined with the relative accuracy of the cross-validation methods lead us to prefer these methods, both latinCV and randomCV, to all current alternative methods.

%
%
%\singlespacing
%\begin{enumerate}
%\item Comparison of Cross-validation methods
%	\begin{itemize}
%	\item NCV appears to perform significantly worse than randomCV in any scenario where the accuracy of either method is above 10-20 percent.
%	\item latinCV appears to have similar accuracy to randomCV.
%	\end{itemize}
%\item Comparison to other methods
%	\begin{itemize}
%	\item AIC tends to perform much worse than the other methods, frequently preferring models that are too complex
%	\item BIC appears to also select more complex models, resulting in increased accuracy when the signal to noise ratio is low.
%	\item Infomap appears to do worse than all of the other methods at performing model selection.
%	\item Modularity Maximization performs relatively well, but is outperformed overall by both BIC and all of the cross-validation methods.
%	\end{itemize}
%\end{enumerate}

%\singlespacing
%
%\clearpage
{
\footnotesize
\bibliographystyle{ims}
\bibliography{cv-method-comparison}

}
\end{document}